\documentclass[fleqn,usenatbib]{mnras}

\usepackage{newtxtext,newtxmath}

\usepackage[T1]{fontenc}

\usepackage{orcidlink}

\DeclareRobustCommand{\VAN}[3]{#2}
\let\VANthebibliography\thebibliography
\def\thebibliography{\DeclareRobustCommand{\VAN}[3]{##3}\VANthebibliography}

\usepackage{graphicx}	
\usepackage{amsmath}	

\title[Continuum Delays in Bright Quasars]{Continuum Reverberation in Bright Quasars Using NASA/ATLAS}

\author[Z. Steyn et al.]
{Zachary Steyn,$^{{\orcidlink{0009-0007-5201-0357}}\,1}$\thanks{E-mail: zachary.steyn@anu.edu.au}
Christian Wolf,$^{{\orcidlink{0000-0002-4569-016X}}\,1,2}$
Christopher Onken,$^{{\orcidlink{0000-0003-0017-349X}}\,1,2}$
Ken Smith,$^{{\orcidlink{0000-0001-9535-3199}}3,4}$
Ji-Jia Tang,$^{{\orcidlink{0000-0002-1860-0886}}\,5,1}$ 
\newauthor
Andjelka B. Kova\v cevi\'c,$^{{\orcidlink{0000-0001-5139-1978}}\,6}$
John Tonry,$^{{\orcidlink{0000-0003-2858-9657}}\,7}$
and Alejandro Clocchiatti$^{{\orcidlink{0000-0003-3068-4258}}\,8,9}$
\\
$^{1}$Research School of Astronomy and Astrophysics, Australian National University, Cotter Road, Weston Creek ACT 2611, Australia \\
$^{2}$Centre for Gravitational Astrophysics (CGA), Australian National University, Building 38 Science Road, Acton ACT 2601, Australia \\
$^{3}$Department of Physics, University of Oxford, Denys Wilkinson Building, Keble Road, Oxford OX1 3RH, UK\\
$^{4}$Astrophysics Research Centre, School of Mathematics and Physics, Queen’s University Belfast, BT7 1NN, UK\\
$^{5}$Research Center for Space and Cosmic Evolution, Ehime University, Matsuyama, Ehime 790-8577, Japan\\
$^{6}$University of Belgrade-Faculty of Mathematics, Department of Astronomy, Studentski trg 16, Belgrade, Serbia\\
$^{7}$Institute for Astronomy, University of Hawaii, 2680 Woodlawn Drive, Honolulu, HI 96822-1897, U.S.A. \\
$^{8}$Millennium Institute of Astrophysics (MAS), Nuncio Monseñor Sotero Sanz 100, Of. 104, Providencia, Santiago, Chile \\
$^{9}$Instituto de Astrofísica, Facultad de Física, Pontificia Universidad Católica de Chile, Av. Vicuña Mackenna 4860, Santiago, Chile
}

\date{Accepted XXX. Received YYY; in original form ZZZ}

\pubyear{\the\year{}}

\begin{document}
\label{firstpage}
\pagerange{\pageref{firstpage}--\pageref{lastpage}}
\maketitle

\begin{abstract}
Variable continuum emission from AGN can be used to probe the structure of their accretion disks via reverberation mapping. Assuming a variable, hot inner light source irradiates the surrounding accretion disk, time delays between different continuum band light curves reveal light-travel times between their respective emission regions. Inter-band delays measured in several low-luminosity AGN are ubiquitously $\sim 3$ times longer than expected from standard disk theory, with evidence this size discrepancy may decrease in more luminous AGN. We have analysed high-cadence light curves of 9,498 of the brightest quasars between redshift 0.3--2.5 in the largest continuum reverberation study to date. Given the large sample size, we construct bins and fit delays jointly to combine inference across the parameter space and improve lag detections. We find that the size discrepancy persists in our high-luminosity sample, and that the previously seen anti-correlation with luminosity is likely driven by wavelength effects. The complex, non-monotonic wavelength dependence of delay amplitudes strongly suggests that contamination of inter-band delays by variable diffuse emission is widespread in the AGN population. We test delay behaviour against a variety of quasar properties finding longer lags in quasars with: higher Eddington ratios, redder colours, stronger optical \ion{Fe}{ii} EWs, higher iron ratios (both UV \ion{Fe}{ii}/\ion{Mg}{ii} and optical \ion{Fe}{ii}/H$\,\beta$), \ion{C}{iv} broad absorption troughs, and lower \ion{C}{iv} blueshift.
\end{abstract}




\section{Introduction}

Investigations of the geometry and kinematics of Active Galactic Nuclei (AGN) are challenged by our inability to resolve the compact sizes of their central regions \citep[for notable exceptions see:][]{2018_gravity_spatially,2019_eht_first}. In lieu of spatial resolution, the wavelength-dependent aperiodic variability and distinct spectral shape of AGN emission can encode information about the underlying physical geometry and processes in said systems. Reverberation mapping \citep[RM;][]{1982_blandford_reverberation,1998_collier_steps} aims to decode this hidden spatial structure by cross-correlating the observed variability of different physical components. The resulting time delays then infer geometric distances provided the coherent variability between each emission region is linked irradiatively.

Although first implemented within the context of the Broad Line Region \citep[BLR;][]{1991_clavel_steps,1993_peterson_reverberation,1997_wanders_steps}, RM has now been used to measure linear sizes for almost all AGN components from X-ray to infrared \citep[see][for a review]{2021_cackett_reverberation}, achieving equivalent microarcsecond spatial resolution. In this work, we focus on continuum RM in AGN accretion disks, where the irradiating source is frequently ascribed to be the X-ray corona. Under this model of a central X-ray 'lamppost', delay times between continuum bands are a measure of the light travel time between the respective emission regions, as the surrounding disk instantaneously reprocesses the incident X-ray stimulus. We thus expect delays to increase with wavelength, as the variable X-ray emission first reaches the hotter inner disk before the cooler outer regions.

Intensive campaigns of local Type 1 AGN with high cadences and wide wavelength coverage report a lag-wavelength dependence consistent with $\tau \propto \lambda^{4/3}$\citep{1998_collier_steps,2017_fausnaugh_reverberation}, as is predicted for the widely adopted non-relativistic thin-disk model \cite[geometrically thin and optically thick;][hereafter SSD for Shakura and Sunyaev Disk]{1973_shakura_black}. The normalisations of lags found in these campaigns are almost universally larger than expected by a factor $\sim 3$ \citep{2016_fausnaugh_space,2018_cackett_accretion,2018_mchardy_x-ray,2019_edelson_first,2021_vincentelli_multiwavelength}. This size discrepancy has been independently reported in microlensing observations of quasar optical emission regions \citep{2007_pooley_x-ray,2010_morgan_quasar,2011_mosquera_microlensing}. 

The literature has generated many potential explanations for the observed size discrepancy, including complex disk geometries \citep{2017_gardner_origin,2023_starkey_rimmed}, internal AGN reddening \citep{2017_gaskell_case}, disk winds \citep{2019_sun_winds,2019_li_reconciling}, and non-blackbody emission \citep{2018_hall_non-blackbody}. We further describe here three popular models that we are poised to examine in our dataset:

\defcitealias{2021_kammoun_uv-optical}{K21}
\begin{itemize}
    
    \item \textbf{Diffuse BLR model:} Diffuse continuum emission from the BLR is expected to respond to changes in the ionising luminosity from the central engine \citep{2001_korista_variable,2018_lawther_quantifying,2019_korista_quantifying}, contributing variable emission across the entire UV-optical spectrum and lengthening delays (as the BLR is more distant than the accretion disk). This introduces a complex wavelength dependence for measured lags (hereafter lag spectra) owing to the wavelength-dependent contribution to continuum bands. Evidence for the BLR model is seen through excess lag signal surrounding the Balmer jump \citep{2016_fausnaugh_space,2018_cackett_accretion}; however, the degree to which this effect is present is not universal across all AGN \citep{2019_edelson_first,2021_kammoun_modelling,2023_kara_uv-optical}. Frequency-resolved lags provide further evidence for a more distant reprocessor as standard cross-correlation delays are dominated by longer timescale variability consistent with the BLR \citep{2022_cackett_frequency,2023_lewin_x-ray}. Diffuse BLR continuum emission may even be the dominant signal in inter-band delays, with BLR models matching several AGN lag spectra with only minor contributions from an irradiated disk \citep{2022_netzer_continuum,2024_netzer_storm}.

    \item \textbf{\citetalias{2021_kammoun_uv-optical} model:} By considering relativistic coronal irradiation, local disk ionization and substantial corona heights, \citet{2021_kammoun_uv-optical} \citepalias{2021_kammoun_uv-optical} find the X-ray reprocessing scenario is consistent with observations, predicting wavelength-dependent lags \citep{2021_kammoun_modelling,2023_kammoun_revisiting}, power spectral densities \citep[PSD;][]{2022_panagiotou_physical}, and spectral energy distributions \citep[SED;][]{2024_kammoun_broadband} in several well-studied AGN. Using the best-fit parameters for NGC 5548 \citep{2025_panagiotou_frequency}, the \citetalias{2021_kammoun_uv-optical} model self-consistently predicts the frequency-resolved lags of \citet{2022_cackett_frequency} on all but the longest timescales where UV-optical variability has been known to exceed the X-ray \citep{2003_uttley_correlated}.

    \item \textbf{CHAR model:} \citet{2020_sun_char} propose a model where the central X-ray corona and the surrounding accretion disk are magnetically coupled, with coronal variability instead thermalised in the disk through the dissipation of the induced magnetic turbulence. Coherent fluctuations propagate from the corona to the disk at the Alfv\'en velocity, with the stochastic disk emission response being damped and delayed on the order of the thermal timescale as the system re-establishes thermal equilibrium. The model is applied to light curves from NGC 5548 \citep{2016_fausnaugh_space}, simultaneously fitting both the measured time delays and the structure functions of all 18 bands. The CHAR model further predicts the frequency-resolved lags of several AGN \citep{2024_chen_frequency,2024_chen_corona-heated}.

\end{itemize}

While the UV-optical emission of AGN shows clear correlated behaviour \citep{1991_clavel_steps,1991_krolik_ultraviolet} with wavelength dependent variability \citep{2004_vanden_berk_ensemble,2014_morganson_measuring} and delay structure \citep{2016_fausnaugh_space,2018_cackett_accretion}, it remains unclear whether X-ray reprocessing is the dominant mechanism driving these phenomena. Some systems show only low-to-moderate correlation between X-ray and UV-optical light curves \citep{2018_buisson_connection,2019_edelson_first,2019_morales_x-ray,2023_cackett_agn}, particularly on short timescales \citep{2024_partington_connecting}. \citet{2022_panagiotou_explaining} suggest this lack of correlation arises naturally when considering a dynamical X-ray corona, where the physical properties and geometric configuration that determine reprocessing are variable, as may be the case in several AGN \citep{2020_caballero-garcia_combined,2020_alston_dynamic,2021_zoghbi_hard,2022_panagiotou_physical}. Strong X-ray-UV-optical correlation is only expected if X-ray emission is sufficiently luminous and disk albedo is sufficiently low \citep{2025_secunda_continuum}. X-ray luminosities can fall well below this threshold, sometimes requiring X-rays to be a factor $\gtrsim10$ more luminous to explain changes in UV-optical emission \citep{2019_dexter_sloan,2023_marculewicz_disk}. Evidence for excess variable emission beyond X-ray reprocessing is predominantly present on longer timescales \citep{2003_uttley_correlated,2009_breedt_long-term,2025_beard_testing}, while short timescale variations can plausibly be accounted for \citep{2022_panagiotou_physical,2024_partington_connecting,2024_papoutsis_x-ray}.

Our picture of coherent AGN variability is further complicated by the detection of inward-travelling intrinsic disk signals. \citet{2020_hernandez_santisteban_intensive} first found evidence of this in Fairall 9, where the low frequency component (determined by a quadratic fit) experienced its minima at earlier times in redder wavelengths. Later, \citet{2023_yao_negative} quantitatively measured Fairall 9's inward travelling lags, positing a steeper radial disk scale height profile as the cause for the expectedly long viscous timescales falling within the observed ranges. Inward travelling fluctuations have also been detected in temperature variations \citep{2022_neustadt_using,2023_stone_temperature,2024_neustadt_storm}, while flux light curves show the commonly seen lamp-post behaviour on short timescales.

While intensive studies have provided a wealth of high quality data to test interesting and complex accretion theories on, other works have foregone the intense monitoring of single objects, exchanging higher wavelength and time resolution for more modest specifications and greatly increased sample size to study a broader population of AGN. These studies typically cover more massive and more luminous AGN with larger anticipated delays, owing to an insufficient observational cadence to resolve lags in the smaller, individually studied AGN. 

In large time-domain survey studies, the presence of a disk size discrepancy is much less universal. \citet{2017_jiang_detecting} present evidence for the 'disk too big' problem in their sample of 240 quasars using Pan-STARRS1 survey data \citep{2012_schlafly_photometric}, although estimates may be biased by making a significance cut which preferentially selects objects with larger lags given the moderate cadence. Using light curves from the Dark Energy Survey \citep[DES;][]{2015_kessler_difference}, \citet{2018_mudd_quasar} derive continuum lag measurements for 15 quasars, finding agreement with both the reported size issue and the standard theory with an elevated (but still sub-Eddington; $L/L_{\rm Edd} < 1$) accretion rate due to their large uncertainties. \citet{2019_homayouni_sloan} take a Bayesian approach to estimating a disk size normalisation parameter for their sample of 95 well-defined lags from the first year of the SDSS-RM campaign \citep{2015_shen_sloan}, with results consistent with SSD predictions within $\pm 1.5\sigma $. \citet{2020_yu_quasar} present accretion disk sizes for 22 quasars in the DES standard star and supernova C fields, finding agreement with SSD predictions if the radii probed are weighted by their response to X-ray stimulus rather than their steady-state flux (this predicts lags a factor $\sim1.5$ larger than the standard luminosity-weighted assumption). 

The larger sample sizes of continuum RM survey studies have enabled correlating lags with a range of physical parameters, giving further clues to the origin of inter-band delays. Meta-analysis combining delays from \citet{2019_homayouni_sloan} and \citet{2020_yu_quasar} with several intensively studied Seyfert galaxies reports an anti-correlation between the ratio of observed-to-SSD-predicted lags with luminosity \citep{2021_li_faint}. The authors find that both BLR contamination \citep[if diffuse BLR continuum equivalent width is anti-correlated with luminosity as is the case for UV broad-lines;][]{1977_baldwin_luminosity} and the CHAR model can explain their results. For a sample of 94 AGN, \citet{2022_guo_active} produce lags from Zwicky Transient Facility \citep[ZTF;][]{2019_masci_zwicky} light curves, reporting the presence of both elevated disk size measurements and an anti-correlation in the observed-to-SSD-predicted lag ratio with luminosity. The estimated radius at 5100\AA\, also follows a tight scatter with the luminosity at 5100\AA\, ($R_{5100}\propto L_{5100}^\alpha$ with $\alpha = 0.48\pm0.04$), suggesting an intrinsic relation between the accretion disk and BLR sizes \citep[which scale similarly with continuum luminosity;][]{2013_bentz_low-luminosity,2021_kaspi_taking,2024_shen_key}. This conclusion is supported by \citet{2023_wang_estimating}, who collate both continuum disk ($R_{5100}$) and BLR (H$\beta$) size estimates for several objects in the literature \citep[half of which come from][]{2022_guo_active}, finding a tight scatter ($\sim$0.28 dex) between the two. \citet{2024_sharp_sloan} further examine the quasars present in \citet{2019_homayouni_sloan}, also finding an anti-correlation of observed-to-SSD-predicted lag ratio with luminosity. However, they do not find evidence of diffuse BLR or \ion{Fe}{ii} emission in the rms spectra and are unable to recreate the tight scatter seen in \citet{2023_wang_estimating} when using SDSS-RM derived H$\beta$ radii \citep{2017_grier_sloan}, instead citing the CHAR model as a more likely explanation.

It is clear that the current state of continuum reverberation mapping is a complex topic of open discussion. To break degeneracies between literature theories, we need to examine lags for a broad population of AGN. We tackle this problem by conducting the largest continuum RM study to date with $\sim 9,500$ of the most UV-optical luminous spectroscopically confirmed type 1 quasars, using $\sim$3-day-cadence photometry from the Asteroid Terrestrial-impact Last Alert System \citep[ATLAS;][]{2018_tonry_atlas}. With this large and luminous sample we hope to verify whether the previously seen anti-correlation between the disk size discrepancy and luminosity persists, and if so, try discern between the BLR, \citetalias{2021_kammoun_uv-optical}, and  CHAR models. 

In Section \ref{sec:data}, we discuss our sample selection process and describe the structure and cleaning of the ATLAS light curves. Section \ref{sec:sims} outlines the calculation of theoretical lags and the generation and verification of simulated light curves for independently testing our ability to obtain expected results with the ATLAS data structure. A description of literature-standard RM delay algorithms and our specific implementation of them is provided in Section \ref{sec:methods}. Results from the ATLAS data are presented in Section \ref{sec:results} with discussion therein. We adopt a flat $\Lambda\mathrm{CDM}$ cosmology with $H_0 =70\, \mathrm{km}\,\mathrm{s}^{-1}\,\mathrm{Mpc}^{-1}$ and $\Omega_\Lambda = 0.7$. All magnitudes presented are in AB units.

\section{Sample and Data}\label{sec:data}

Reverberation mapping replaces spatial resolution with temporal resolution, and like any telescope, there is a minimum necessary resolution to probe certain structures. While there is no concrete agreement on how fine temporal sampling must be in irregularly sampled light curves, successful campaigns usually have a finer cadence than the expected delay signal \citep{2021_cackett_reverberation}. Continuum RM at its minimum requires light curves in two passbands that probe different rest-frame UV-optical emission regions in the AGN's accretion disk. While not strictly necessary, an increased wavelength coverage is desirable to compare delays between similar physical radii across redshifts and to investigate the radial profile that sets wavelength dependency.

\subsection{Quasar Samples}

\subsubsection{Main Sample Selection}\label{sec:main_sample}

\defcitealias{2020_rakshit_spectral}{R20}

\begin{figure}
    \centering
    \includegraphics[width=\columnwidth]{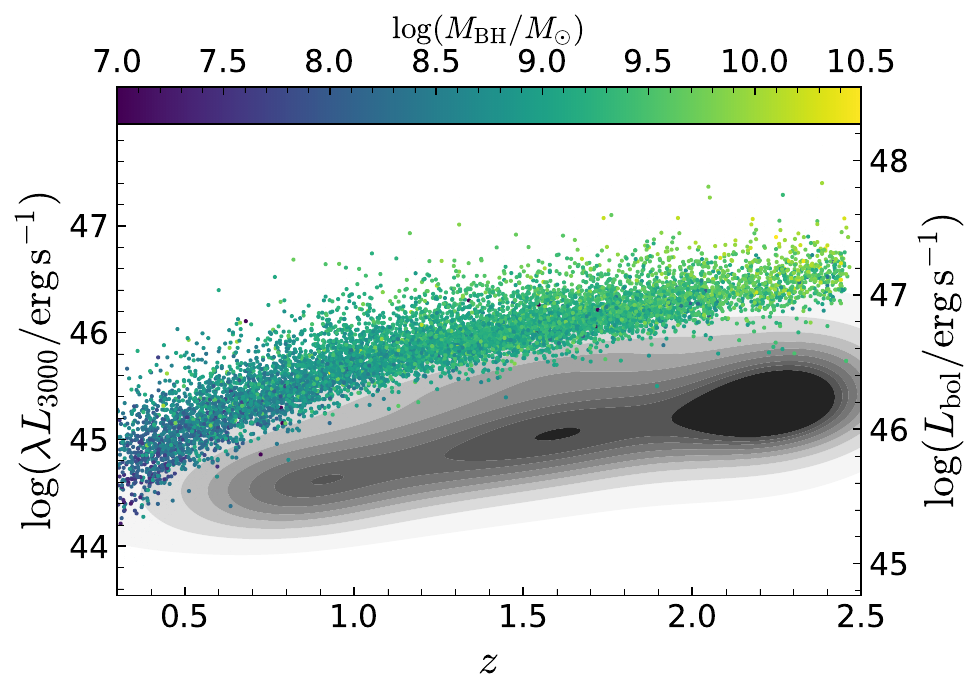}
    \caption{Parent sample: luminosities and black hole masses vs. redshift compared to the wider SDSS quasar population illustrating our chosen sample is far among the most luminous. Bolometric luminosities are estimated from the \citetalias{2020_rakshit_spectral} $\lambda L_{3000}$ values using a bolometric correction of 5.15 \citep{2006_richards_spectral}. A weaker correlation of black hole mass with both quantities is present.}
    \label{fig:sample_stats}
\end{figure}

We construct our parent sample by selecting all spectroscopically confirmed type-1 quasars in the SDSS DR14 quasar catalogue \citep[][hereafter \citetalias{2020_rakshit_spectral}]{2020_rakshit_spectral} that lie within the ATLAS field of view (see Section \ref{sec:ATLAS_lcs}). From here, we retain quasars with good quality spectral decomposition estimates of the $\lambda L_{3000}$ continuum available. This ensures that we have the best estimates of accretion disk luminosity for our SSD size predictions, free from iron continuum and line emission contamination. The sample is further restricted to quasars that are sufficiently bright (Gaia magnitude $R_p < 18$), unobscured by Milky Way dust ($E(B\,-\,V)_{\mathrm{SFD}}<0.1$), and persistently low airmass (Dec $<+70$), retaining quasars with the highest average signal-to-noise in both ATLAS bands. This leaves a sample of 12,055 quasars with bolometric luminosities between $10^{45}$--$10^{48}{\rm\, erg\,s}^{-1}$ that are strongly correlated with redshift (Figure \ref{fig:sample_stats}), as is expected in a flux limited sample.

We reject quasars with strong radio counterparts to avoid potential contaminating variability from jet processes or non-standard disk structure. Following \citet{1989_kellerman_vla}, objects with a ratio of radio-to-optical flux ({\it R}; 6\,cm to $B$ band) greater than 100 are classified as radio loud. The original sample of \citet{1989_kellerman_vla} sits at a median redshift $z \sim 0.5$, corresponding to approximate rest-frame wavelengths of 4\,cm and 3000\AA\, for radio and optical emission respectively. The optical luminosity ($L_{3000}$) is taken from the \citetalias{2020_rakshit_spectral} catalogue. We estimate radio emission by cross-matching with the {\it NRAO VLA Sky Survey} \citep[NVSS;][]{1998_condon_vla} and the {\it Faint Images of the Radio Sky at Twenty-centimeters} survey \citep[FIRST;][]{1995_becker_first}, taking the brighter 1.4\,GHz flux of the two within 30 arcsec. We then extrapolate a rest-frame 4\,cm luminosity assuming a mean radio spectral slope of $-0.3$ \citep{1988_kellermann_galactic}. This slope choice is conservative for steep-spectrum quasars at $z\lesssim 4$, cautiously rejecting a greater number of sources. We make no further attempts to distinguish between radio-quiet ($R<3$) and radio-intermediate ($3<R<100$) as NVSS and FIRST are not sensitive enough to make a complete distinction in our sample. This leaves a remaining 10,950 quasars.

\subsubsection{Removing Objects with Contaminating Variability}\label{sec:remove_exc_var}

We remove objects with prominent variability from external sources that would dilute the disk RM signal. This includes variable PSF contamination from bright neighbours due to atmospheric seeing and possible microlensing events. As quasars have characteristic variability properties, we can examine objects that greatly depart from the norm to identify candidates for removal. Characterisation and comparison of quasar variability is performed on the cleaned ATLAS light curve data (refer to Section \ref{sec:ATLAS_lcs} for details).

To prevent contamination from bright nearby sources, objects are required to be isolated within a 15" radius in the Gaia catalogue. To test whether this criterion is sufficient, we compare the ATLAS light curve rms values of quasars with bright nearby stars (Gaia $R_p$<12.5, <10.0, <7.5, and <5.0 mag within 30", 60", 120" and 240" respectively) to the distribution of rms values for similar quasars (in $\lambda L_{3000}$ and $z$) without bright nearby sources. We find no bias in rms values for quasars with bright nearby sources, finding rms values spread uniformly across the comparison rms cumulative distribution function (CDF) independent of angular separation. This suggests that the ATLAS photometry is uncontaminated for our remaining sample.

To search for potential microlensing events, we visually inspect all 10,950 light curve pairs. We find and remove two conspicuous candidate microlensing events based on this visual inspection, leaving the detailed analysis to a future paper. Due to the stochastic nature of AGN emission, weak microlensing contamination can be degenerate with intrinsic fluctuations, even by eye. Thus, we only remove objects where microlensing contributions appear significant. These candidate microlensing events are seen in quasars with the highest fractional changes in flux emission (90th percentile and above) for their respective luminosity. We further remove 11 objects whose light curve pairs show notably strange behaviour or questionable quality.

\subsubsection{Removing the Dust-reddened Tail}\label{sec:remove_dust}

Quasars reddened by host galaxy dust will have fainter continuum luminosity estimates, and thus smaller anticipated accretion disk sizes \citep{2017_gaskell_case,2023_gaskell_estimating}. Quasars that are systematically reddened compared to the widespread population are thus removed from our main sample. To determine this, we derived median $c-o$ colour estimates from each object's cleaned NASA/ATLAS light curve pair (see Section \ref{sec:ATLAS_lcs}) using a linear interpolation of a 50-day running median to obtain time coincident magnitude estimations for all epochs. To correct for Galactic extinction, we follow \citetalias{2020_rakshit_spectral}, choosing the \citet{1998_schlegel_maps} dust map and the Milky Way extinction law of \citet{1999_fitzpatrick_correcting} with $R_V=3.1$. This ensures that derived colours are as indicative of uncorrected dust extinction in the \citet{2020_rakshit_spectral} $\lambda L_{3000}$ estimates as possible. 

\begin{figure}
    \centering
    \includegraphics[width=\columnwidth]{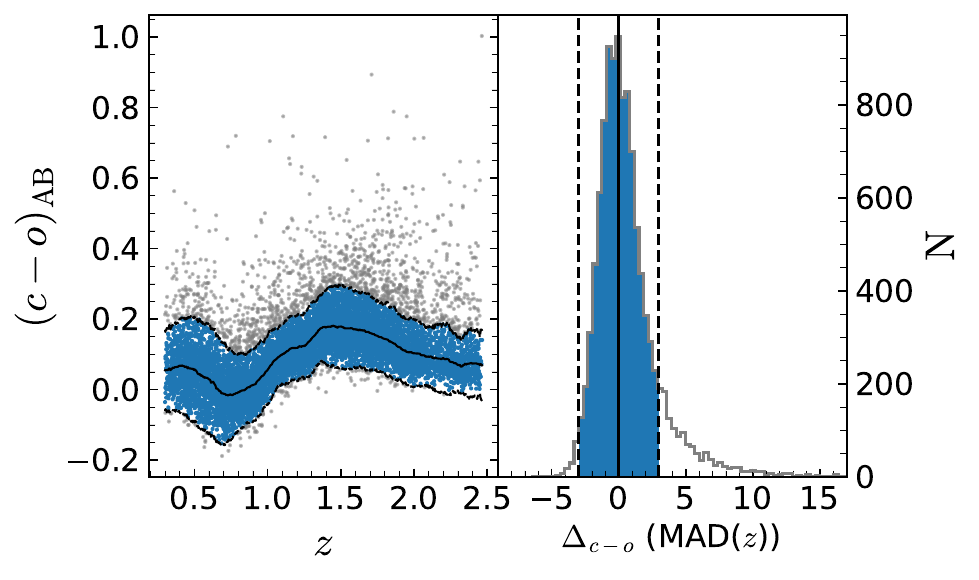}
    \caption{Left: ATLAS colour dependence on redshift. Right: Distribution of colour residuals standardised by redshift-dependent median and MAD. The median ($\pm3$MAD) is shown as the solid (dashed) black line in both panels. Grey points and grey distribution tails represent removed objects.}
    \label{fig:reddened_quasars}
\end{figure}

The corrected median $c-o$ colour for each object shows a strong dependence on redshift as prominent line emission migrates through our filters (Figure \ref{fig:reddened_quasars}). Thus, we fit a running median and median absolute deviation (MAD) with width $\Delta z =0.1$ and remove objects whose median colour lies outside $\pm3$MAD from the running median. This process is repeated a second time as the proportion of dust reddened quasars was large enough to influence the initial running median and MAD. Two iterations were sufficient to produce a convergent, near-symmetric distribution of median colour that contains our main quasar sample of 9,498 objects. Of those removed, 1304 objects belonged to dust-reddened tail \citep[a similar fraction to that seen in][]{2003_richards_red} and 135 objects exhibited substantially bluer colour for their corresponding redshift. 

\subsubsection{\ion{Fe}{ii} Subsamples}\label{sec:fe_sample}

We consider the role of iron emission may play in signalling changes in continuum RM behaviour. While not typically considered in the BLR contamination framework due to its weaker equivalent width \citep[EW;][]{2022_homayouni_sloan} and difficulty to characterise through modelling \citep{2022_netzer_continuum}, the presence of \ion{Fe}{ii} emission is expected to increase measured delays \citep{2024_netzer_storm} and potentially provide a better description of Fairall 9 lags when interpreted with an additional iron BLR reprocessor \citep{2024_edelson_intensive}. We also wish to explore delay behaviour on $R_{\rm Fe}=$ \ion{Fe}{ii}/H$\,\beta$ (and its UV analogue \ion{Fe}{ii}/\ion{Mg}{ii}) as one of the principal variables by which quasar spectral diversity can be explained \citep{2000_sulentic_eigenvector,2007_sulentic_eigenvector}. The latter of these two measures may also serve as a metallicity indicator \citep{2011_de_rosa_evidence}. We thus define subsamples partitioned by \ion{Fe}{ii} EW and the ratio of \ion{Fe}{ii} EW to a prominent low-ionisation line EW (\ion{Mg}{ii} or H$\,\beta$) to search for potential delay dependences.

\begin{figure}
    \centering
    \includegraphics[width=\linewidth]{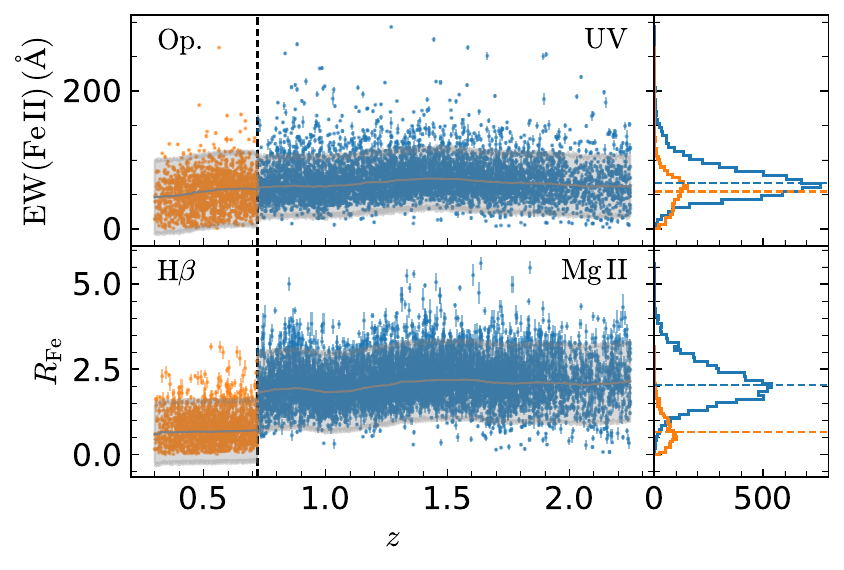}
    \caption{Redshift evolution of \ion{Fe}{ii} EW and its ratio to the relevant low-ionisation line in each regime (separated by the black dashed line $z=0.72$). A running median and MAD trend is displayed in grey. The marginal distributions for each regime are given in the right-hand panels, along with the distribution medians.}
    \label{fig:fe_redshift_evo}
\end{figure}

As our sample covers a significant redshift range, we are unable to compare a single iron complex across the entire sample. We take \ion{Fe}{ii} estimates from the UV (optical) complex above (below) a redshift of $z=0.72$ (panels 7 and above in later binning). Similarly, we choose \ion{Mg}{ii} (H$\,\beta$) for our low-ionisation line above (below) this value. This boundary achieves the greatest number of valid estimates (EW $>0$) while aligning with our binning strategy (see Section \ref{sec:bin_stack}). To be confident in the amount of iron emission in our sample, we require a median signal-to-noise per pixel >10 around the relevant 3000\AA/5100\AA\, continuum. We also require uncertainties in UV \ion{Fe}{ii} EW, optical \ion{Fe}{ii} EW, UV \ion{Fe}{ii}-to-\ion{Mg}{ii} ratio, and optical \ion{Fe}{ii}-to-H$\,\beta$ ratio to be less than 10\AA, 6\AA, 0.4, and 0.25 respectively. Figure \ref{fig:fe_redshift_evo} shows the redshift evolution and overall distribution of the iron emission derived quantities. We see no evidence of significant redshift evolution with these parameters, motivating a single value cut-off to separate our subsamples. This value is chosen the median given the approximately symmetric, unimodal distributions. This leaves a remaining 4140 (4138) quasars with low (high) \ion{Fe}{ii} EW and 4074 (4074) quasars with low (high) iron ratios.

\subsubsection{Outflow Subsample}\label{sec:civ_sample}

\begin{figure}
    \centering
    \includegraphics[width=\columnwidth]{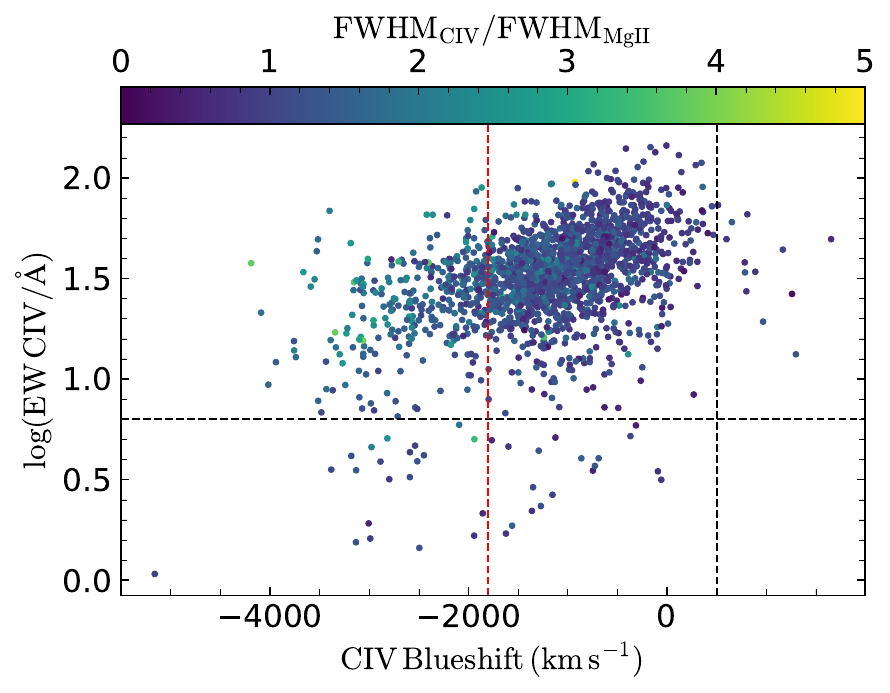}
    \caption{\ion{C}{iv} blueshift anti-correlation with equivalent width for the non-BAL quasars with sufficient line SNR, luminosity and FWHM. Black dashed lines represent sample cuts for equivalent width and blueshift. Red dashed line represent the $\leq-1800\,{\rm km\,s^{-1}}$ wind velocity cut-off.}
    \label{fig:civ_blueshift_EW}
\end{figure}

As we extend the redshift range probed from previous studies, the high-ionisation \ion{C}{iv} line becomes visible in the SDSS spectra, with \ion{C}{iv} blueshift serving as a diagnostic for strong wind outflows \citep{2004_leighly_hubble,2011_richards_unification}. We can thus investigate whether quasars with strong winds exhibit different lag behaviour. 
To ensure \ion{C}{iv} line properties can be accurately determined, we require a median S/N ratio >15 around the \ion{C}{iv} (1500-1600\AA) and \ion{Mg}{ii} (2700-2900\AA) lines, with both line luminosity and FWHM being greater than zero. Objects that show broad UV absorption line features (BAL\_FLAG$\geq 1$) are also removed. We further require lines to be sufficiently prominent $(\log(\rm EW_{\rm \ion{C}{iv}}/A)\geq 0.8$ and not excessively broad $(\rm FWHM_{\rm \ion{C}{iv}}/\rm FWHM_{\rm \ion{Mg}{ii}} \leq 3.5)$ to be confident fitted line properties represent real line structure (these criterion are visualised in Figure \ref{fig:civ_blueshift_EW}). We calculate \ion{C}{iv} blueshift as the blueshift of the peak \ion{C}{iv} emission relative to the blueshift of the peak \ion{Mg}{ii} emission \citep[as \ion{Mg}{ii} is the best indicator of the systematic redshift available in the SDSS spectral range at our redshifts;][]{2016_shen_velocity}.

As we wish to test if the presence of a strong outflow affects lag measurements, we divide our subsample into two bins. Quasars with \ion{C}{iv} blueshifts $\leq-1800\,{\rm km\,s^{-1}}$ are labelled as having strong \ion{C}{iv} outflows. This threshold is chosen as quasars with these outflow velocities have systematically lower variability amplitudes compared to the general quasar population \citep{2023_tang_probing}, although similar behaviour is seen when separating at $-550\,{\rm km\,s^{-1}}$ \citep{2018_sun_civ}. Objects with peak \ion{C}{iv} emission redshifted more than 500km/s relative to \ion{Mg}{ii} velocities are removed from the subsample as these may be a sign of an intrinsically different phenomenon rather than having 'no wind'. This leaves a remaining 1512 (347) quasars with low (high) \ion{C}{iv} blueshift.

\subsection{ATLAS Light Curves}\label{sec:ATLAS_lcs}

ATLAS was developed in partnership between NASA and the University of Hawaii to identify near-Earth and potentially hazardous asteroids (NEAs and PHAs respectively) through the use of time domain observations. Beginning its program in June 2015, ATLAS started with a single 0.5\,m telescope at Haleakala (ATLAS-HKO) in Hawaii. Shortly after, in February 2017, a second identical telescope was added at Mauna Loa (ATLAS-MLO), also in Hawaii. Both the ATLAS-HKO and ATLAS-MLO units can view a declination range of $-45^\circ < \delta < +90^\circ$ which, after accounting for the unobservable sky within $60^\circ$ of the Sun, leaves an accessible sky of about 24,500 deg$^2$ on any given night. With a pixel scale of 1.86 arcsec and a 10560x10560 CCD, each telescope can capture 28.9 deg$^2$ of sky in a single exposure (excluding the edge mosaic). This means a single telescope can image the entire accessible sky in less than 900 exposures, or at 30s per exposure (+10s for readout and re-pointing), within one night. The installation of an additional two units at Sutherland Observatory in South Africa and El Sauce Observatory in Chile in early 2022 has further increased ATLAS coverage in the southern hemisphere and improved the median cadence.

The ATLAS system uses two broadband filters, cyan ($c$: $4157-6556$\AA) and orange ($o$: $5582-8249$\AA), with a limiting $5\sigma$ magnitude of 19.8 in both bands. Although much broader than the typical filters used in continuum RM, the observed frame wavelengths probed are similar to the widely used $gri$ filters \citep{2017_jiang_detecting,2018_mudd_quasar,2019_homayouni_sloan,2020_yu_quasar,2022_guo_active}, with approximate conversions given by $c\sim 0.49\,g\, +\,0.51\,r $ and $o\sim0.55\,r\,+\,0.45\,i$ \citep{2018_tonry_atlas}. The ATLAS system also utilises a unique asymmetric cadence structure between the bands with cyan measurements taken only during dark sky periods. Orange band observations are taken irrespective of lunar phase.

\subsubsection{Bespoke ATLAS Dataset}

This work uses a bespoke dataset created by the ATLAS team from their proprietary data products. Light curves are constructed by first obtaining astrometric and photometric measurements for sources within every ATLAS science exposure using a specially modified version of \textsc{dophot} \citep{1993_schechter_dophot,2012_alonso-garcia_uncloaking}. For a detailed description of the custom \textsc{dophot} implementation and its version-specific modifications see Sect.~2 of \citet{2018_heinze_catalogue}. Summarised briefly, a smooth, spatially varying PSF model of modified elliptical Gaussians (appropriate for point sources like quasars) is fit first to the brightest sources. Bright sources are progressively subtracted and the PSF model is refit to allow for better determination and measurement of fainter objects. Significant detections in each image, and the associated telescope boresight pointing, are catalogued allowing us to iteratively crossmatch and aggregate any measurements within 3.6" ($\sim$2 ATLAS pixels) of our target positions.

The resulting light curves differ from those available on the ATLAS Forced Photometry Server \footnote{Available at \url{https://fallingstar-data.com/forcedphot/}}, which instead uses the specialised ATLAS photometry software \textsc{tphot} to re-measure the photometry at the user-provided coordinates. \textsc{tphot} achieves superior performance when measuring photometry from differenced images and is ideal for extracting the light curves of variable sources, such as supernovae, that coincide with extended static sources. However, photometry produced by \textsc{tphot} from raw images is susceptible to contamination from neighbouring objects and variations in the background flux. As information about the relative flux variability and total emission differences between passbands can provide important context for continuum RM, we preference using \textsc{dophot} to obtain photometry from unsubtracted science exposures.

\subsubsection{Cleaning Individual Measurements}\label{sec:cleanATLAS}

To clean the ATLAS light curves, we remove individual photometric observations that are likely not representative of the true flux. Observations taken before June 2017 are excluded due to degraded image quality, with field corrector lenses replaced in May 2017. This reduces our light curve length to a typical 8--9 seasons per object. We remove observations around the edge mosaic of the detector (within 100 pixels of the x or y boundary). Measurements that deviate from point source behaviour (not classified as either image type 1: point source or type 7: undetermined) or show large disagreement between aperture photometry and PSF fit (${\rm abs(AP-FIT) < 1.5}$) are removed. Images with unusual zeropoints ($m_{\rm zeropoint}\notin[25,26]$), poor seeing (PSF major axis > $4^"$), and high background brightness (cyan and orange sky counts > 1000 and 3000 respectively) are also removed. We exclude observations with relatively large errors (more than twice the median error in the light curve of a given object). Any remaining outlying measurements are determined by subtracting a running median with a 50-day width and removing observations outside $\pm4$ median absolute deviations ($\pm2.7\sigma$ for a Gaussian distribution).

\subsubsection{Aggregating Observations Within A Night}\label{sec:lc_agg}

As the core function of the ATLAS project is to track and constrain trajectories for moving objects, ATLAS reduces its nightly coverage to image each field (up to) four times over the course of an hour with an individual telescope. Given the steep drop off in AGN variability amplitude towards higher frequencies \citep{2011_mushotzky_kepler,2018_smith_kepler}, we expect to see no meaningful variability over the course of an hour relative to the flux uncertainties, and thus choose to aggregate measurements over these one-hour windows through inverse variance weighting. We do not aggregate measurements between telescopes in the same night as observing conditions on any given night may be dissimilar. Light curve sampling properties after aggregating within a night are show in Table \ref{tab:lc_sampling}.

\begin{table}
    \centering
    \begin{tabular}{c c c c c}
         Filter & $\Delta t_{\rm median}$ & $\Delta t_{\rm min}$ & $N_{\rm epoch}$ & $(\sigma_f/f)_{\rm median} $\\
         \hline
         $c$ & $4.02^{15.96}_{2.92}$ & $1.86^{3.97}_{0.94}$ & $15^{26}_{9}$ & $0.03^{0.04}_{0.02}$ \\
         \hline
         $o$ & $2.96^{3.98}_{1.65}$ & $0.88^{1.73}_{0.01}$ & $56^{84}_{43}$ & $0.04^{0.05}_{0.03}$
    \end{tabular}
    \caption{Median per-season values with 16th and 84th percentiles. $\Delta t$ values are provided in observed frame in units of days.}
    \label{tab:lc_sampling}
\end{table}

As having fewer than four observations per night could be an indicator of degraded viewing conditions (the presence of structured cloud), we wish to determine whether a minimum number of observations for aggregation is required. Such conditions may mean that our uncertainties are underestimated, presenting as non-physical variability that dilutes the desired RM signal. We compare having one, two, and three observations per night to our closest source of ground truth: four observations in an aggregated epoch. The 'ground truth' fluxes for comparison are derived from linear interpolation of a running median with a width of 50 days. The difference in flux between aggregating $<4$ observations per night and the estimated 'ground truth' is normalised by the associated flux uncertainty to compile a set of standardised residuals across a random 10\% subsample of light curves. We do not find strong evidence of underestimated flux uncertainties with reduced $\chi^2$ values falling between 0.9--1.6 and 0.8--1.1 for the cyan and orange bands respectively. The slight elevation in reduced $\chi^2$ is more likely due to the imperfect imputation strategy employed that is insufficient to mimic the intrinsic short timescale variability present.

Using this same framework, we perform a test to identify the presence of any large discrepancies between data from different telescopes. Unlike comparing the number of observations per night, there is not an obvious choice of which telescope closest resembles the underlying truth. Given our interpolation strategy, we instead opt to use the telescope with the largest number of observations in each filter, with this being ATLAS-MLO for cyan and ATLAS-HKO for orange. We do not find strong evidence for a measurement offset between telescopes with the mean of each residual distribution consistent with zero. We do note larger reduced $\chi^2$ values in this test (between 0.9--2.6 for cyan and 1.0--1.6 for orange) but again argue this is due to an imperfect imputation method that is sensitive to time-sampling density and separation as, unlike the previous comparison, observations with the lowest flux uncertainties (four-observations-per-night) are no longer necessitated to be time coincident with the input 'ground truth' points. This point is further supported by considering that the input points to the 'ground truth' approximation achieve the lowest reduced $\chi^2$ in both bands and that the cyan 'ground truth' has approximately half the number of contributing epochs as for orange. Even if small biases remain present, we argue this would have a minor effect measuring lags given the statistical nature of our analysis where epochs are all considered, rather than isolating a particular feature.

\section{Simulations}\label{sec:sims}

Cadence structure and observed photometric noise have a complex relationship with resolving RM lags. To verify that we can recover the expected X-ray reprocessing signal in the ATLAS light curves, we simulate comparable light curves for which we know the underlying truth. As we have observational data to anchor our simulated light curves, we need only predict a lag for each object, and then generate a light curve that mimics the expected variability structure. We note here that we use the model defined assuming X-ray irradiation, but it is easily generalised to assume a different, centrally emitted wavelength like the EUV/FUV, which is more strongly correlated with the variable UV-optical emission \citep{2024_partington_connecting}.

\subsection{Theoretical Lags}\label{sub:lag_theory}

With several competing literature models available, we default to the SSD model as our null hypothesis. We refer back to our introduction where we motivated this paper, in part, by aiming to explore whether the anti-correlation of the ratio of observed-to-expected lags with luminosity extends to the most luminous quasars. Under the SSD model, intrinsic disk viscosity facilitates the outward transport of angular momentum, converting gravitational potential energy into heat as the material is accreted inwards. Assuming Keplarian dynamics, a massless disk, and local blackbody emission, the effective temperature profile of the disk due to viscous heating is given by \citep{1973_shakura_black},

\begin{equation}
    T_\mathrm{visc}^4(R) = \frac{3GM_\mathrm{BH}\dot{M}}{8\pi\sigma R^3} \left[1-\left(\frac{R_{\mathrm{ISCO}}}{R} \right)^{1/2} \right],
\end{equation}

\noindent where $R_{\mathrm{ISCO}}$ is the radius of the innermost stable circular orbit (ISCO; 6 gravitational radii $R_g$ for a non-rotating black hole), $\sigma$ ($G$) is the Stefan-Boltzmann (gravitational) constant, and $M_{\rm BH}$ and $\dot{M}$ are the mass and accretion rate of the black hole, respectively. 

The lamppost model assumes the surrounding disk is irradiated by an external X-ray source. While the precise geometry of this 'lamppost' is unknown, it is commonplace in continuum RM to approximate it as a point source sitting at height $H$ above the central black hole. This simplification is appropriate as \citet{2017_gardner_origin} show there is little difference between this approximation and considering a radially extended corona. The irradiative local temperature contribution at each annulus for a flat disk is given by \citep{2016_starkey_accretion},

\begin{equation}
    T_\mathrm{irr}^4(R) = \frac{L_X(1-A)H}{4\pi\sigma(H^2 + R^2)^{3/2}}, 
\end{equation}

\noindent where $L_X$ is the luminosity of the X-ray lamppost and $A$ is the albedo of the disk (assumed to be radially independent).
For the combined local temperature, we include a GR correction factor assuming a face-on disk inclination to account for gravitational redshift and time-dilation effects in the observed temperature profile \citep{1989_hanawa_x-ray},

\begin{equation}\label{eq:T_GR}
    T_{\mathrm{GR}}(R) = \sqrt{1-\frac{3GM_\mathrm{BH}}{Rc^2}}\left[T_\mathrm{visc}^4(R) + T_\mathrm{irr}^4(R)\right]^{1/4}. 
\end{equation}

\noindent As the radii discussed here in the X-ray reprocessing scenario are far larger than the black hole's ergosphere, we do not correct for Lense-Thirring effects. Relativistic beaming is also neglected as this effect is more dominant in edge-on viewing scenarios not expected with the near face-on assumption of type-1 quasar geometry.

To estimate expected delays across our disks, translate our observed SDSS parameters of mass and continuum luminosity into a mass-accretion rate. Given a black hole mass estimate, we apply an iterative solver to determine mass-accretion rate such that the observed monochromatic rest-frame luminosity $\lambda L_{3000}$ is recovered using

\begin{equation}
    L_\lambda = \int^{R_\mathrm{out}}_{R_\mathrm{isco}}\frac{4 \pi hc^2 \cos{i}}{\lambda^5} \frac{R \,dR}{e^{hc/\lambda k T(R)}-1},
\end{equation}

\noindent where $h$ and $k$ are the Planck and Boltzmann coefficients respectively, and the radial temperature profile $T(R)$ is given by Eq \ref{eq:T_GR}. An average viewing inclination of $i\sim 30^{\circ}$ \citep{1989_barthel_quasar} is assumed to account for the isotropy assumption used in \citetalias{2020_rakshit_spectral}. For objects with poor black hole mass quality flags, we instead use the median black hole mass of quasars within $\pm0.1$ dex $\lambda L_{3000}$ and $\pm0.1$ redshift. This approximation is appropriate as black hole mass causes only a second order effect on expected delays (see Figure \ref{fig:lag_ratio} and surrounding discussion). Temperature profiles are evaluated logarithmically in radius, using Eq \ref{eq:T_GR} with the interior radii set to $6R_g$. The exterior radii are determined by the $R\gg R_{\mathrm{ISCO}}$ limit ($T \propto R^{-3/4}$) such that $T(R_{\rm out}) \approx 1500K$, which is a typical dust sublimation temperature for silicate grains \citep{1987_barvainis_dust}. An estimate for the lamppost luminosity is calculated assuming that the monochromatic X-ray luminosity can be described by a photon index $\Gamma=2$ over the 2--10 keV range with $\log(L_{2\rm keV})=(0.71\pm 0.03)\log(L_{2500})+(5.0\pm 1.0)$ \citep{2021_liu_observational}. Here, the $L_{2500}$ is derived from $\lambda L_{3000}$ assuming a near ultraviolet (NUV) slope of $\alpha=-0.3$ \citep[$f_\nu \propto \nu^\alpha$;][]{2016_xie_luminosity}. The lamppost is assumed to be irradiating from a height $H = 10R_{\rm g}$, as the X-ray emitting region is measured to be in the vicinity of the ISCO \citep{2012_morgan_further,2014_blackburne_optical}. The albedo is set to zero, creating an upper bound for the lamppost contribution, although estimates are not sensitive to the particular value $A\in[0,1)$ given the temperature profile in large disks is dominated by internal viscous heating for this model. 

\begin{figure}
    \centering
    \includegraphics[width=\columnwidth]{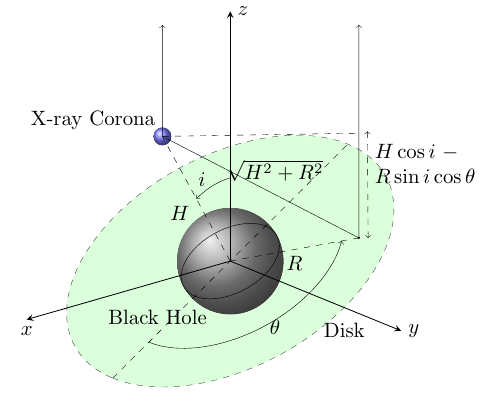}
    \caption{Time delay seen by a distant observer along the $z$-axis for an X-ray photon reprocessed by the disk where $i$ is the system inclination and $R$ ($\theta$) is the radial (azimuthal) position.}
    \label{fig:lag_diagram}
\end{figure}

\begin{figure}
    \centering
    \includegraphics[width=\columnwidth]{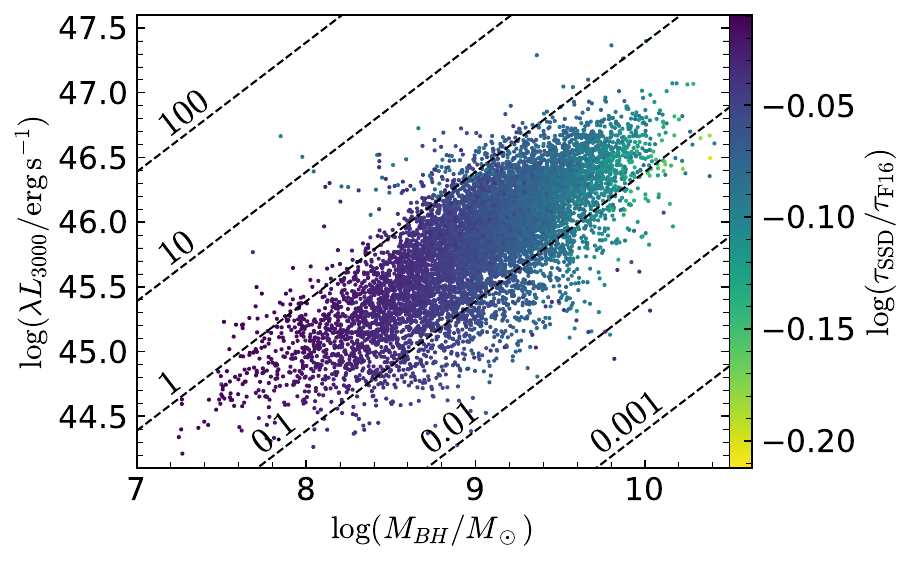}
    \caption{Ratio of expected delays derived here to those estimated from Eq. 9 of \citet{2016_fausnaugh_space} for our main quasar sample in the $\lambda L_{3000}$-$M_{\rm BH}$ plane. Black dashed lines represent quasars of fixed Eddington ratio $\dot{m}_{\rm edd}$}.
    \label{fig:lag_ratio}
\end{figure}

Reverberation mapping frequently assumes a linear relationship between driving $D(t)$ and response $X(t)$ lights curves, as given by

\begin{equation}\label{eq:linearRM}
    X_\lambda(t)=\int\psi(\tau|\lambda)D(t-\tau)d\tau
\end{equation}

\noindent where $\psi(\tau|\lambda)$ is the transfer function and is given by \citep{1998_collier_measuring}

\begin{equation}
\psi(\tau|\lambda)=\int^{2\pi}_0\int^{R_\mathrm{out}}_{R_\mathrm{isco}}\frac{\partial B_\nu}{\partial T}\frac{\partial T}{\partial L_X}\frac{R\,dR\, d\theta\cos{i}}{D^2}\times \delta\left(\tau - \tau(R,i,\theta)\right),
\end{equation}

\noindent where $B_\nu$ is the Planck function, $D$ is the distance to the AGN, and $c\tau(R,i,\theta)=\sqrt{H^2+R^2}+H\cos{i}-R\sin{i}\cos{\theta}$ (see Figure \ref{fig:lag_diagram}). The expected lag between bands is given as difference between their respective transfer function centroids. Here, we evaluate the transfer functions for the effective wavelength of each filter, finding little difference when accounting for the finite filter widths. We also note a $\lesssim 0.06$ dex difference comparing with $c\tau(R,i,\theta)=R(1+\sin{i}\cos{\theta})$, with deviation occurring primarily when $M_{\rm BH} \gtrsim 10^9M_\odot$. We do not include the first-order GR correction in the temperature profile when calculating the transfer function, with the responsivity of the disk requiring a more careful GR consideration not given here \citepalias{2021_kammoun_uv-optical}.

As accretion rate is somewhat degenerate with black hole mass when constrained only by continuum luminosity, we test whether black hole mass provides strong independent information in the SSD delay paradigm. We compare our expected lags to those derived using Eq. 9 of \citet{2016_fausnaugh_space} using $X = 5.04^{3/4}$ for converting temperature to radius assuming radii are weighted by their response to the incident radiation \citep{2018_tie_microlensing}. This analytical prescription does not include a dependence on the ISCO (and hence black hole mass), instead treating $M_{\rm BH}\dot{M}$ as a scaling parameter (constrained only by luminosity). Figure \ref{fig:lag_ratio} shows the ratio of the two methods has a weak dependence on black hole mass, with the inclusion of the ISCO geometry deviating by less than 0.1 dex from pure scaling behaviour for all bar the highest black hole masses. This value decreases to 0.05 dex if the corona instead sits at $20R_g$. This motivates continuum luminosity and redshift as the main driving parameters in the SSD X-ray reprocessing theory, setting the scale of, and the location probed, on the disk, respectively.

\subsection{Generating Simulated Light Curves}\label{sec:lc_gen}

With the above transfer function machinery, we can generate simulated light curves for each of the ATLAS bands from a given X-ray light curve. We assume the null hypothesis that X-ray reprocessing is the dominant variability mechanism and choose to model X-ray light curves as a random walk (${\rm PSD} \propto f^{-2}$). Under linear reprocessing, this ensures the corresponding UV-optical light curves match the observed random walk variability  seen on rest-frame timescales of 10--250 days in a subset of our current quasar sample using ATLAS data \citep{2024_tang_variability}. Generating light curves through transfer function convolution suppresses short timescale variability below the random walk assumption, especially for the larger disks. Such suppression is not inconsistent with results from \citet{2024_tang_variability}, although the more simple interpretation of a pure random walk is favoured.

With a specified PSD shape, we use the \citet{1995_timmer_generating} algorithm to generate a random X-ray light curve to convolve with the relevant transfer functions for each passband. Light curves are generated to have an observed frame sampling of 0.1 days, and have a length at least 10 times the observational baseline to account for red-noise leakage \citep[the tendency for light curves to show power on timescales longer than the observed baseline;][]{2002_uttley_broadband}. Quasars with similar luminosity and redshift to our sample have shown random walk behaviour may extend out to at least the long timescales assumed here \citep{2022_stone_optical}. Power on such long timescales is inconsistent with X-ray reprocessing in real AGN as X-rays exhibit flatter PSD slopes on long timescales \citep{2023_paolillo_universal,2024_prokhorenko_x-ray,2024_georgakakis_ensemble}, but their inclusion is necessary for calibrating against the real ATLAS data. We do later compare light curve behaviour with and without this long timescale variability through detrending.

\begin{figure}
    \centering
    \includegraphics[width=\columnwidth]{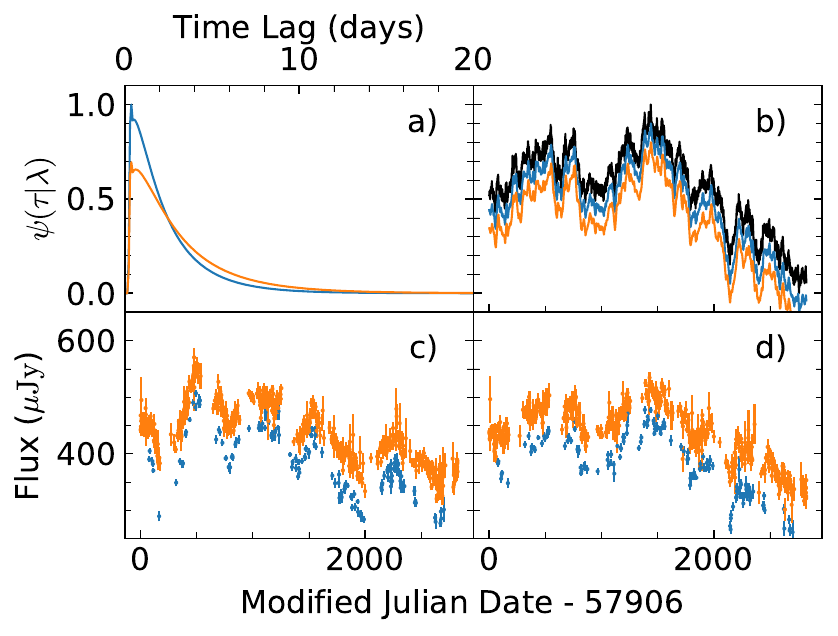}
    \caption{(a) Example transfer function with arbitrary normalisation evaluated for an object with $\log(\lambda L_{3000}/{\rm erg\, s}^{-1})=45.02$, $\log(M_{\rm BH}/M_\odot)=8.47$, $z=0.309$, and $\dot{M}=0.53M_\odot\,{\rm yr}^{-1}$. (b) Generated X-ray light curve (black) with associated cyan and orange fluxes. Flux values and offsets are arbitrary and chosen for display purposes. (c) Observed ATLAS light curves for the example object. (d) Calibrated simulated light curve based on panels b and c. Both panel c and d apply a $-50\mu{\rm Jy}$ offset to cyan epochs for clarity.}
    \label{fig:sim_lc_example}
\end{figure}

To calibrate the generated UV-optical light curves we sample epochs through linear interpolation to match the observed sampling distribution before normalising to the mean and variance of the real light curves. Each simulated epoch is given the corresponding observed flux uncertainty and is permuted with Gaussian noise accordingly. The resulting set of light curves have near identical fractional variability to their observed counterparts with known delay and variability structure as specified by standard X-ray reprocessing theory. This process is illustrated for an example object in Figure \ref{fig:sim_lc_example}. 

\subsection{Inter-band Correlation Properties}

We further verify whether our simulations mimic the true ATLAS data by comparing the peak inter-band correlation ($r_{\rm max}$ -- calculated using ICCF; see Section \ref{sec:iccf}) between the two light curve sets. We make these comparisons on two different timescales: on light curves with and without detrending using independent quadratic fits in each band. As sample cuts based on observed light curve correlation are frequently made within the literature \citep{2019_homayouni_sloan,2022_guo_active}, we also examine whether the inter-band correlation of UV-optical emission may be related to the expected lag magnitude.

\begin{figure}
    \centering
    \includegraphics[width=\linewidth]{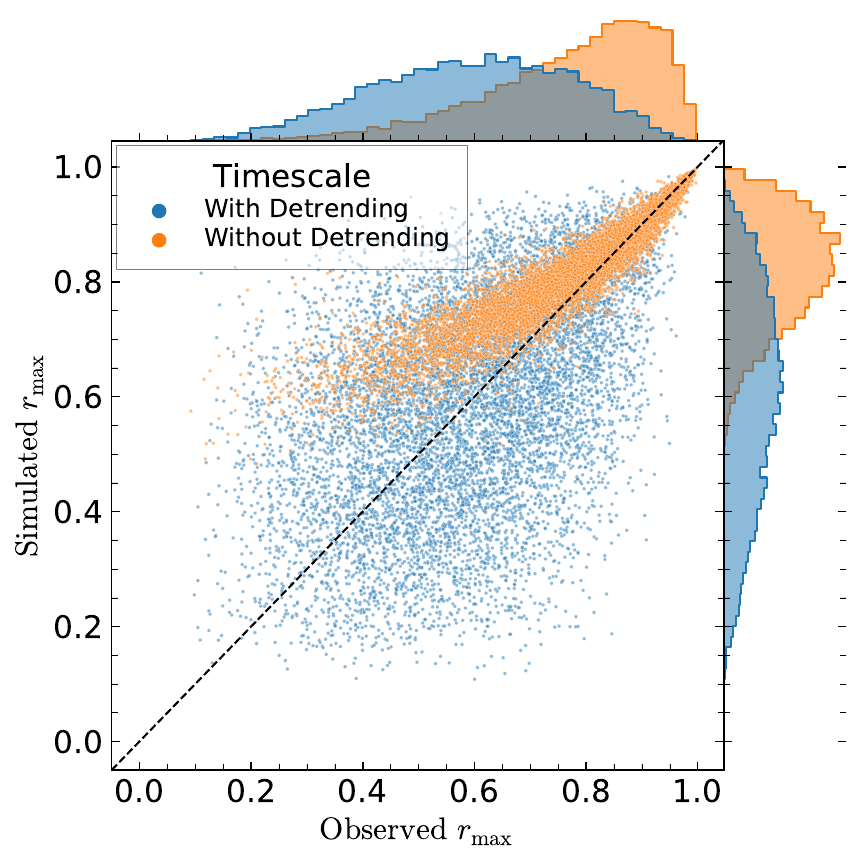}
    \caption{Pairwise distribution of ICCF-derived $r_{\rm max}$ values in the observed and simulated datasets. Values are shown for light curves both with (blue) and without (orange) quadratic detrending. Associated marginal distributions are shown as well.}
    \label{fig:rs_comparison}
\end{figure}

In the detrended light curves, the pairwise distribution of inter-band correlations is largely symmetric about the one-to-one line with dispersion increasing towards lower $r_{\rm max}$ values (Figure \ref{fig:rs_comparison}). With the strong anti-correlation between $r_{\rm max}$ and the median cyan/orange relative flux uncertainty, the observed diversity and dispersion is plausibly explained by our photometric precision. Light curves with small flux uncertainties will show correlated behaviour irrespective of strong momentary stochastic variability. As quadratic detrending depends on the realised variability and does not filter particular frequencies directly, the inter-band correlation in noisier light curves is strongly subject to how much of the total attributed variance (normalising on fractional variability) remains after detrending. We also note the real ATLAS data having a higher concentration of objects with $r_{\rm max} \in [0.4,0.7]$ and the simulated data having a higher concentration of objects in the $r_{\rm max}<0.4$ tail. At face value, this is unexpected as real AGN light curves may contain some level of non-correlated variability not associated with continuum reverberation, whereas our simulated light curves are intrinsically correlated. Normalising simulated light curves to have near identical fractional variability preserves only the total amount of variability relative to the noise, but not on which timescales that variability is attributed. Given we are comparing short timescale behaviour in the detrended light curves, it is likely the lower average inter-band correlation in our simulated dataset comes from overestimating the power of very long timescale variability. Additionally, the increased inter-band correlation in the real data may be influenced to the partial overlap of the ATLAS filters, whereas the simulated data considers the distinct effective wavelengths only. At this stage we have no reason to suspect either the intrinsic correlations of UV-optical variability differ significantly from those simulated here on short timescales or that this variety of correlation cannot be explained by light curve properties, plausibly allowing for sample cuts based on $r_{\rm max}$ (see further discussion in Section \ref{sec:bin_stack}).

When comparing the longer timescale behaviour of the original cleaned light curves, we see a much smaller level of intrinsic scatter between the simulated and real sets (Figure \ref{fig:rs_comparison}). This is expected as without detrending, inter-band correlation is less dependent on the realised variability and is more dependent on the light curve properties (i.e. fractional variability and cadence) which are conserved. We do note however that the inter-band correlation in this simulated data is higher than the real with the fraction of light curve pairs with a peak correlation $r_{\rm max}<0.6$ being $1.51\%$ and $14.5\%$ respectively. This bias towards stronger inter-band correlation in the simulated data is seen most strongly in objects with lower luminosities and at higher redshifts. As it remains unclear whether this is an unexplained dependence on magnitude or a genuine departure from correlated behaviour, we analyse light curves both with and without detrending (see Appendix \ref{sec:pipeline_check} for comparison).

\section{Estimating Time Delays}\label{sec:methods}

Most common RM algorithms in the literature aim to measure the centroid of the transfer function, either by calculating the cross-correlation function (CCF) empirically (e.g. ICCF) or by assuming a functional form for the transfer function (e.g. \textsc{javelin}). All methods considered here assume a static geometry (lag behaviour is not a function of time), the relationship between light curves is linear (Eq. \ref{eq:linearRM}), and that variability is wide-sense stationary (variability structure is not a function of time).

\subsection{ICCF}\label{sec:iccf}

The interpolated cross-correlation function \citep[ICCF;][]{1998_peterson_uncertainties} attempts to solve our unevenly sampled light curve problem in a pairwise fashion, projecting sampled times from the reference light curve (with an added delay) onto the response. The response light curve is then linearly interpolated to estimate the flux that corresponds to the projected reference times. The Pearson correlation coefficient can then be applied, iterating over a number of delays to construct the CCF, the centroid of which ($\tau_{\rm cent}$) is ideally equivalent to the centroid of the transfer function \citep{1991_koratkar_structure} and is thus taken as our delay estimate. Uncertainties of $\tau_{\rm cent}$ are built up over many Monte Carlo (MC) simulations using random subset selection (RSS; sampling epochs with replacement) and flux randomisation (FR; permuting fluxes by their uncertainties) during each MC realisation. We use the Cython implementation of ICCF \footnote{Available at \url{https://bitbucket.org/cgrier/python_ccf_code}} developed by \citet{2018_sun_pyccf}, which reduces flux uncertainties by a factor $n^{1/2}$ for each time the associated epochs are selected during RSS \citep{1999_welsh_reliability}. Our final lag estimate is taken as the median of the resulting cross-correlation centroid distribution (CCCD), with errors given by the 16th and 84th percentiles.

We default to using detrended light curves in our ICCF analysis, finding this preprocessing is a necessary step for recovering delays, reflecting previous recommendations \citep{1999_welsh_reliability}. We choose to globally detrend using an independent quadratic fit for each filter, as is common within the reverberation mapping sphere \citep{2014_mchardy_swift,2016_fausnaugh_space,2020_hernandez_santisteban_intensive, 2024_edelson_intensive}. We implement the default two-interpolation strategy which calculates the CCF as the average of the individual CCFs produced when each passband is treated as the interpolated response curve. We do not alter the default thresholds for peak correlation significance ($r_{\rm max}\geq0.2$; also the lower bound of our short timescale $r_{\rm max}$ distribution) or the boundaries for calculating the CCF centroid value ($r>0.8r_{\rm max}$). CCFs are calculated across $\pm100$ days in 1 day increments, building up 10,000 MC samples using FR+RSS. We see no difference in results when extending the search range to $\pm200$ days. We also do not consider any anti-aliasing weighting of the CCF \citep[see][for examples]{2017_grier_sloan,2020_yang_dust} as the range of lags explored is far less than the length of our light curves. Such methods become important when exploring delays with minimal overlapping points between light curves by down-weighting correlations that arise spuriously from the low sample counts.

\subsection{\textsc{JAVELIN}}\label{sec:javelin}

\textsc{javelin} \footnote{Available at \url{https://github.com/nye17/javelin}} \citep{2011_zu_alternative} is an RM algorithm that models noisy light curves with a Gaussian process \citep{1992_rybicki_interpolation}. Light curve values $\mathbf{y} = \mathbf{s} + \mathbf{n} + L\mathbf{q}$ are represented as a combination of inherent signal $\mathbf{s}$, observational noise $\mathbf{n}$, and a general trend $L\mathbf{q}$ (by default the mean flux value in each passband) and are comprised of both the driving ($D(t)$) and response ($X(t)$) curve observations. Entries of the signal covariance matrix $S=\left< \mathbf{ss}\right>$ are calculated assuming the driving curve follows a damped random walk (DRW),

\begin{equation*}
    \left <D(t_i),D(t_j) \right> = \sigma_{\rm DRW}^2 e^{-|t_i - t_j|/\tau_{\rm DRW}},
\end{equation*}

\noindent where $\sigma_{\rm DRW}$ is the variability amplitude on short timescales and $\tau_{\rm DRW}$ is the decorrelation timescale. The covariance between observations in different bands, or within the response band, can then be calculated through appropriate convolution of the DRW kernel and the transfer function as the light curves, and hence the covariances, are assumed to be linearly related (Eq. \ref{eq:linearRM}). \textsc{javelin} assumes the true transfer function can be represented by a tophat function with parameters $A$, $\tau$, and $w$ representing the scale, temporal shift (lag), and smoothing width respectively. 

By assuming Gaussian likelihoods for the signal, noise, and linear terms, and marginalising over $\mathbf{s}$ and $\mathbf{q}$, the likelihood of observing the data given the DRW kernel and tophat function parameters $\mathbf{p}$ is evaluated as,

\begin{equation}\label{eq:jav_likelihood}
    P(\mathbf{y}|\mathbf{p}) \propto \mathcal{L} = | S + N |^{-1/2} |L^T C^{-1} L |^{-1/2} \mathrm{ exp}\left ( -\frac{\mathbf{y}^TC_\perp^{-1} \mathbf{y}}{2}\right ),
\end{equation}

\noindent where $N$ is the covariance matrix of the noise terms, $C = S + N$, and $C_\perp^{-1}$ is the component of $C$ orthogonal to the fitted linear parameters $\mathbf{q}$. Eq. \ref{eq:jav_likelihood} is passed to a Markov Chain Monte Carlo (MCMC) sampler to build up posteriors of model parameters, of which the 16th, 50th, and 84th percentiles are taken as our lower bound, estimate, and upper bound. In practice, DRW priors are first obtained by running \textsc{javelin} on just the reference curve.

Our \textsc{javelin} analysis uses light curves without detrending (see Appendix \ref{sec:pipeline_check} discussion), finding delays are better constrained without this preprocessing step. \textsc{javelin} is also prone to aliasing \citep{2016_fausnaugh_space,2020_yu_reverberation} and will return spurious peaks at lags that prevent overlap between light curves as there is no data to reject the proposed parameters. For this reason, outlier exclusion is essential, particularly for those with small uncertainties, as \textsc{javelin} will preference aligning outliers present in both light curves, or preference placing the outlier in observational gaps if present in only one light curve. Given the seasonal gaps present in our data, we restrict lag searches to be between $\pm50$ days to prevent aliasing. We find little evidence that lags outside of this window are present in our sample (as shown by ICCF centroids). Unlike ICCF's centroid calculations, \textsc{javelin} does not rely on neighbouring points to estimate lags and thus this smaller search width is chosen.

\subsection{Binning \& Stacking}\label{sec:bin_stack}

It is common for large, survey-based continuum RM studies to derive robust lags for only a fraction of their candidate light curves (see Table \ref{tab:prev_RM_counts}). Given that measuring a delay depends heavily on observed light curve properties such as sampling rate, flux uncertainty, and momentary stochastic variability, it is reasonable to assume that we are still in a regime limited by our data capabilities rather than probing two classes of AGN, with and without inherent time delays. Previously, light curves that could not constrain a delay are discarded. We describe here a method to extract the information present in these discarded light curves to improve the measurement of the underlying continuum RM signals present in our dataset.

\begin{table}
    \centering
    \begin{tabular}{l|c|c|c|l}
        Paper & $N_{\rm cand}$ & $N_{\rm lags}$ & $\Delta t_{\rm median}$ & Data Source \\
        \hline
        This work & \multicolumn{2}{c}{9\,498} & $\sim 3-4$ & ATLAS  \\
        \hline
        \citet{2017_jiang_detecting} & 240 & 102 & $\sim$3 & PanSTARRS1\\
        \citet{2018_mudd_quasar} & 771 & 15* & $\sim$7 & DES\\
        \citet{2019_homayouni_sloan} & 222 & 95(33) & $\sim$2.5 & SDSS RM\\
        \citet{2020_yu_quasar} & 802 & 22 & $\sim$1 & DES\\
        \citet{2022_guo_active} & 455 & 94(38) & $\sim 3-4$ & ZTF\\
        \hline
    \end{tabular}
    \caption{Number of candidate objects versus the number of successful lags for different cadences (in units of days) and data sources from previous survey continuum RM studies. Bracketed counts are for a more stringent success definition. *\citet{2018_mudd_quasar} select based on variability, not lag measurements.}
    \label{tab:prev_RM_counts}
\end{table}

In order to accurately constrain lag estimates as a function of our sampled parameter space, we combine inference through stacking. This allows us to marginalise over influential physical properties and states that are not well constrained (orientation, line/diffuse continuum emission, spin, etc). This includes time-dependent phenomena such as disk-wind column density \citep{2021_kara_storm,2024_lewin_storm}, coronal properties \citep{2020_caballero-garcia_combined,2020_alston_dynamic,2022_panagiotou_physical}, and momentary stochastic variability. Stacking is already implicit when light curves are analysed as a whole, as delays themselves can be time-varying \citep{2024_lewin_storm}. Declination effects such as season length, airmass-dependent average seeing, and photometric quality are also mitigated. The remaining lags are then presented for the 'average' AGN as a function of known parameters.

Evidence of complex wavelength dependence in lag spectra \citep{2016_fausnaugh_space,2018_cackett_accretion} prioritises binning resolution in wavelength over luminosity -- which has a monotonic relation with delay magnitude \citep{2022_guo_active,2023_wang_estimating}. We thus divide our sample space into 10 uniform bins of $1/(1+z)$ before further dividing the space in 5 uniform bins of $\log \lambda L_{3000}$. After conditioning on $1/(1+z)$, each luminosity bin sees scatter of $\pm0.2$ dex in $\lambda L_{3000}$. As discussed in \ref{sub:lag_theory}, we do not control for black hole mass as its effect is subdominant for the majority of our sample in the SSD prescription. We also test a further 28 $1/(1+z)$ bins shifted one-quarter out of phase to verify trends are independent of Poisson noise from binning choices, although only the independent bins are presented throughout.

Stacking increases the amount of discriminatory information that may be lacking for a single object approach. To stack lag information we take an approach similar to that of previous BLR RM work \citep{2013_fine_stacked,2017_li_sloan,2024_malik_ozdes} by combining information at the inference level. Specifically, for ICCF we leverage the Pearson cross-correlation coefficient's equivalent representation as the mean product of standard scores,

\begin{equation}
    r_{xy}(\tau) = \frac{1}{N(\tau)-1}\sum^{N(\tau)}_{i=1}\left ( \frac{x_{t_i+\tau}-\bar{x}}{s_x}\right)\left( \frac{y_{t_i}-\bar{y}}{s_y}\right).
\end{equation}

\noindent where $x_{t_i}$ ($y_{t_i}$), $\bar{x}$ ($\bar{y}$), and $\sigma_x$ ($\sigma_y$) are the the flux at time $t_i$, mean flux, and standard deviation of each light curve respectively. With this representation, we can combine CCF estimates between similar quasars as
\begin{equation}
\begin{split}
    r_{\mathrm{stack}}(\tau) &= \frac{\sum^M_{j=1}(N_j(\tau)-1)\,r_{j}(\tau)}{\sum^M_{j=1}N_j(\tau) - 1} \\
    &=\frac{1}{\sum^M_{j=1}N_j(\tau)-1}\sum^M_{j=1}\sum^{N_j(\tau)}_{i=1}\left ( \frac{x_{t_i+\tau,j}-\bar{x}_j}{s_{x,j}}\right)\left( \frac{y_{t_i,j}-\bar{y}_j}{s_{y,j}}\right),   
\end{split}
\end{equation}
\noindent where $i$ indexes the observations for each quasar and $j$ indexes each quasar in the bin. This construction is equivalent to having a much longer observational baseline of an individual 'class' of quasar disks while enforcing a level of stationarity through de-trending (see Section \ref{sec:iccf}) and preventing interpolation between the constituent objects. Such an approach has the added benefit of addressing limitations in the ICCF centroid measurement due to the finite sampling of the underlying autocorrelation function \citep{1999_welsh_reliability}. We do not choose to stack by season as well as by object, as the large anticipated lags in our disks require computing cross-correlations at lags that greatly diminish the number of overlapping epochs in a single season. 

\begin{figure*}
    \centering
    \includegraphics[width=\textwidth]{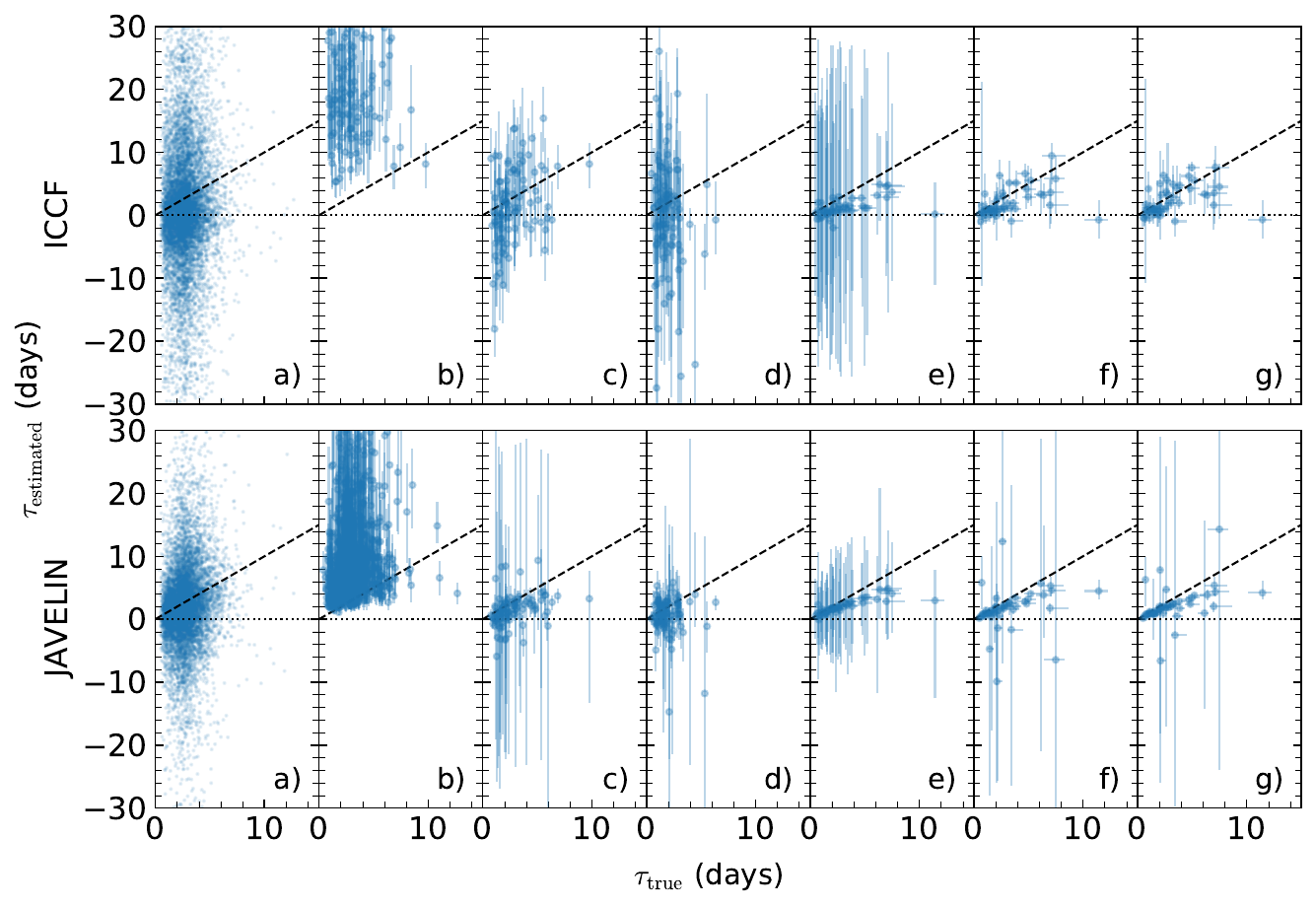}
    \caption{(a) We apply ICCF (top) and \textsc{javelin} (bottom) to each simulated light curve pair in our sample individually, plotting the recovered versus simulated underlying delay. Errorbars are not displayed for clarity. (b) Similar to plot a), but restricting to recovered lags to be >0 at the 1-$\sigma$ level and have $|(\tau_{84}-\tau_{16})/\tau_{50}|<1$. (c) Similar to a), but requiring detrended light curve inter-band correlation to be in the top 1\%. (d) Similar to a), but keeping only objects with the 1\% most variable objects based on detrended orange light curves. (e) Binned posterior samples from the individual applications to each constituent object. (f) Identical binning to e), instead applying the stacked ICCF approach. (g) Identical to f), requiring the constituent objects to have a detrended light curve inter-band correlation > 0.5.}
    \label{fig:sim_comparison}
\end{figure*}

To verify whether stacking CCFs offers improved lag detection capability, we apply the method to the detrended dataset simulated in Section \ref{sec:sims}. Figure \ref{fig:sim_comparison} (top) compares the stacked approach to several other implementations of ICCF, illustrating the following: Panel A shows that with the available quality of our light curves, we are unable to accurately constrain the underlying lag by applying ICCF to light curve pairs individually. For Panel B, if we apply quality cuts -- requiring positive lags, $\geq1\sigma$ significance, and low percentage errors ($|(\tau_{84}-\tau_{16})/\tau_{50}|<1$) -- we preferentially select longer lags and bias our recovered lag distribution. Panel C instead restricts lags based on detrended light curve inter-band correlation, showing a distribution of lags that appears approximately unbiased, albeit with lower significance at the short delay end. Panel D makes a similar cut to Panel C, instead based on detrended orange band variability (measured as $\sqrt{S^2-\left<\sigma^2\right>}$ where $S^2$ is the light curve variance and $\left< \sigma^2\right>$ is mean squared flux uncertainty). The orange band is chosen to probe the underlying short timescale variability as it is much more densely sampled than cyan. A similar effect is seen in the variability-selected lag distribution, as short timescale inter-band correlation and variability probe similar principles: the amount of signal present relative to the noise. In Panel E, combining posterior samples also alleviates the bias, but does little to further constrain the inherent lag distribution except for the longest lags. Panel F shows that to more accurately constrain the true lag distribution, we must stack at the inference level i.e. with cross-correlation functions. Slight improvement is again offered in Panel G as we restrict the constituent light curves to have detrended inter-band correlation greater than 0.5. This threshold is chosen as it retains $\sim2/3$ of the sample, keeping sample statistics high. The resulting distribution appears to achieve slightly better performance for short delays when compared to Panel F stacking.

We can apply an analogous method to stacking with \textsc{javelin} as we did with ICCF. Under the reasonable assumption that epochs in each object are independent from every other object, we can combine the likelihood functions of all objects within the bin to infer a joint parameter set

\begin{equation}
    \log \mathcal{L}_{\rm stack} = \sum_{j=1}^M \log \mathcal{L}_j.
\end{equation}

Quasars with similar luminosity and redshift will have similar variability- ($\sigma_{\rm DRW}$ and $\tau_{\rm DRW}$) and transfer function parameters ($A$, $\tau$ and $w$) and thus can be estimated jointly. Any dispersion in estimating $\sigma_{\rm DRW}$ and $\tau_{\rm DRW}$ due to the range of observed momentary variability is not expected to influence results as the importance of these parameters for recovering lags is minor \citep{2020_yu_reverberation}.

We conduct near-identical tests on the simulated light curves without detrending to verify the benefit of stacking \textsc{javelin} (Figure \ref{fig:sim_comparison}; bottom). The behaviour seen here broadly mirrors that of the ICCF comparisons. Bar a few outliers, the core distribution of uncertainties produced by \textsc{javelin} are consistently smaller than for ICCF, as is noted in several other studies \citep[e.g.][see also Appendix \ref{sec:delay_props}]{2019_edelson_first, 2020_yu_reverberation}. Interestingly, \textsc{javelin} individual applications performed better when cut by short-timescale inter-band correlation and fractional variability even though \textsc{javelin} is applied to light curves without detrending.

\begin{figure*}
    \centering
    \includegraphics[width=\textwidth]{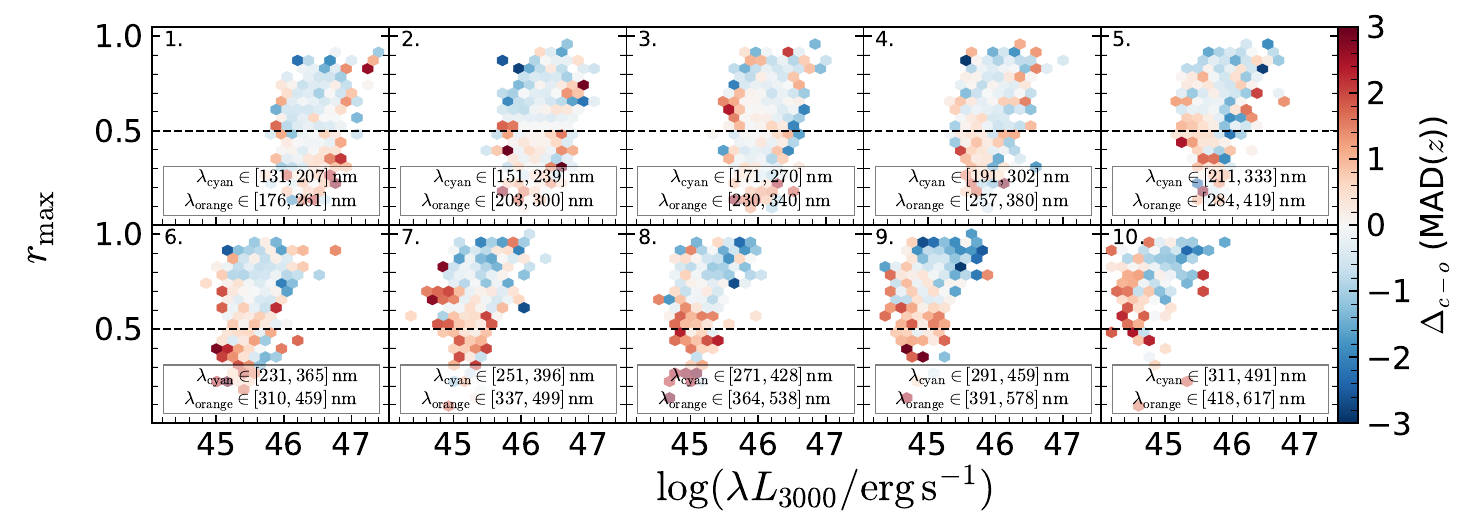}
    \caption{Short-timescale inter-band correlation versus luminosity for each the representative set of independent $1/(1+z)$ bins (with displayed wavelengths given by the filter edges at the $1/(1+z)$ midpoint). Each hexagon is coloured by the median colour residual in units of the redshift dependent spread (MAD) shown in Figure \ref{fig:reddened_quasars}. Bluer quasars on average have stronger short-timescale inter-band correlation. The black dashed line represents the $r_{\rm max}\geq0.5$ cut imposed for the stacking.}
    \label{fig:col_res_hexbin}
\end{figure*}

Despite the (albeit minor) improvement in recovering simulated lags when cutting by short-timescale inter-band correlation, we preference the full stacked approach used in panels F of Figure \ref{fig:sim_comparison} for our further analysis. Figure \ref{fig:col_res_hexbin} shows that the short-timescale inter-band correlation is correlated with luminosity at fixed redshift which is expected as brighter quasars will have smaller photometric uncertainties on average. Requiring that $r_{\rm max}\geq0.5$ thus asymmetrically removes a larger number of quasars from the lower luminosity bins. Figure \ref{fig:col_res_hexbin} also displays the median colour residual relative to the redshift trend identified in Figure \ref{fig:reddened_quasars} across the available parameter space. We see that cutting by inter-band correlation will also preferentially make bins bluer on average by removing red sources (particularly in low-luminosity bins) and thus altering our final results through a colour dependent effect (see Section \ref{sec:colour_lags}). We note that the correlation-dependent behaviour seen here, and in the following discussion, is also reflected in the long-timescale inter-band correlation (which is not shown for brevity).

To investigate whether the link between short-timescale inter-band correlation with quasar residual colour is driven by cyan light curve quality (the limiting band), we match blue quasars ($\Delta_{c-o}\in[-3,-1)$) and red quasars ($\Delta_{c-o}\in[1,3]$) based on their nearest median cyan relative flux error ($\sigma_c/f_c$) counterpart with a maximum difference of 0.005. Quasars can be matched to multiple colour counterparts but duplicate matches are dropped. We divide this matched sample into the previously described $1/(1+z)$ bins and confirm with a one-sided z-test that the median orange percent flux error ($\sigma_o/f_o$) is larger among the bluer quasars when controlling for $\sigma_c/f_c$ as expected. We repeat this test now for short-timescale inter-band correlation and find that redder quasars still have significantly lower correlations (in all bar panels 3/4) even with comparable cyan, and more precise orange, light curves. This suggests that lower inter-band correlation is an intrinsic trait of redder quasars rather than a data driven property. Interestingly, we see similar behaviour (albeit with smaller test statistics) in the simulated dataset. As the simulated data is fundamentally correlated by design with identical sampling and flux uncertainties as the real light curves, this dependence must be encoded through the variability amplitude.

We test whether the fractional variability of the original light curves (to which our simulations are anchored) differs significantly with residual quasar colour. The fractional variability is significantly lower in redder quasars in both the cyan and orange (except panel 4). This is especially surprising for the orange passband where the relative errors are smaller in redder quasars (with comparable cyan errors to bluer counterparts). This behaviour is consistent with dust reddening if the redder quasars are intrinsically brighter and thus less variable \citep{2009_kelly_variations,2016_simm_panstarrs,2024_chanchaiworawit_ensemble}. Alternatively, the reduced variability could be explained by a higher Eddington ratio \citep{2010_macleod_modelling,2013_kelly_active,2016_kozlowski_revisiting} as we find a correlation between residual colour and Eddington ratio in our sample. In spite of controlling for relative flux error (i.e. magnitude for a uniform survey exposure time) and redshift (i.e. distance), we can still see a statistical difference in luminosity for some bins, with redder quasars appearing more (less) luminous in high (low) redshift bins. We retest binning by both $1/(1+z)$ and $\lambda L_{3000}$ as described for the delay analysis above. Of the 50 independent bins we use to compare, we drop 15 for having fewer than 10 unique quasar pairs. Of the remaining 35 bins, 19 bins have statistically less well correlated red quasars and 13 bins have statistically less correlation and fractional variability while not having significantly more luminosity. There does not appear to be a pattern across the parameter space with significance (bar panels 3/4 containing no significant bins).

While stacking delay estimates at the inference level offers clear utility in constraining the underlying delay distribution, it suffers from a lack of flexibility. Delay experiments must be chosen prior to analysis. Testing for delay dependence on any new physical parameterisation requires re-binning and stacked lags to be re-calculated. This becomes computationally expensive on the large datasets where the stacked methodology is most applicable. Individual lags only need to be derived once and correlations can be drawn easily thereafter and are thus preferenced when data quality allows (unlike here; see Figure \ref{fig:sim_comparison}).

\subsection{Multi-peaked Behaviour}\label{sec:multi-peak}

Delay distributions possessing multiple peaks is a known issue in AGN time delay analysis. These multi-modal distributions can be produced in the presence of semi-repeating features, particular cadence structures, or poor sample statistics when exploring delays that cause minimal overlap between light curves. The delay distributions in this study are no exception. We wish to identify which measurements may contain artificial signal not associated with real reverberation.

A common approach to remove signal aliases is to smooth delay distributions with a Gaussian kernel before identifying a primary peak and adjacent troughs. A weighting scheme can also be applied to down-weight potentially spurious signals at lags that cause minimal overlap. New lag estimates can then be derived from the truncated distribution subject to possessing a minimum probability mass \citep{2017_grier_sloan,2019_homayouni_sloan,2020_yang_dust}. 

Given our use of a stacked delay analysis, it is entirely plausible to observed multiple peaks in the recovered lag distributions -- particularly in bins with low counts where the inherent distribution may not be unimodal. Disentangling which peaks are artificial and which are real becomes difficult in all but the most obvious cases (many peaks separated at regular intervals) without a complete catalogue of predictors that can influence delays on the order of separations seen.  As such, we choose not to remove bins that display multiple peaks in their lag distributions, but instead reduce the opacity (i.e. visual weighting) in all further plots. This way, bins that observe potentially real diverse lag behaviour are not discounted, but are not given the same standing as bins that are perhaps better marginalised and representative of the average quasar in that parameter space. The number of peaks here is classified from the number of local maximums in a smoothed posterior lag distribution that exceed 10\% of the peak density. Smoothed distributions are obtained through convolution with a Gaussian kernel with a two-day width. We provide discussion on what light curve properties influence the detection of multiple posterior peaks for each delay algorithm in Appendix \ref{sec:delay_props}.

\section{Results and Discussion}\label{sec:results}

\subsection{Main Sample}\label{sec:main_results}

Armed with the relevant tools to predict and measure inter-band continuum delays in our sample, we can now explore what properties influence lags in real AGN. We begin our search in the main sample, where we have the largest number statistics available and the best chance of constraining robust lag measurements. This sample also has the widest coverage in luminosity and redshift available to us. One of the defining goals of this work was to extend the dynamic range of AGN luminosities for which continuum delay measurements are available, and thus explore whether the anti-correlation of observed-to-expected delays with continuum luminosity persists in our highly luminous sample. 

\begin{figure*}
    \centering
    \includegraphics[width=\textwidth]{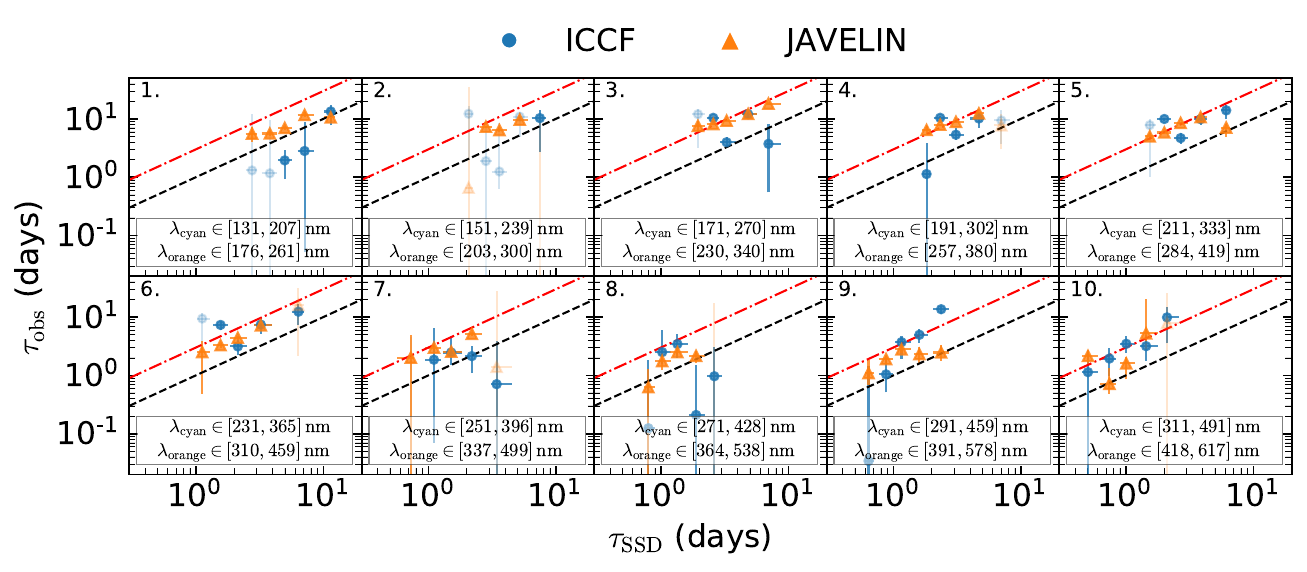}
    \caption{Observed frame measured delays (symbols) vs. SSD expectation (dashed lines are SSD sizes, red dash-dotted lines are $3\times \tau_{\rm SSD}$) for each luminosity bin at fixed redshift range (panels are list in order of $1/(1+z)$) . The displayed wavelength ranges are the filter edges for the bin's $1/(1+z)$ midpoint. Low opacity indicates delay posteriors with multiple peaks (Section \ref{sec:multi-peak}).}
    \label{fig:main_results_lag}
\end{figure*}

Given the inherent correlation between luminosity and redshift in our flux-limited sample, we choose to explore lag dependence on luminosity while controlling for rest-frame properties. Figure \ref{fig:main_results_lag} provides this comparison for the binning strategy described in Section \ref{sec:bin_stack}. Here, luminosity is translated into units of the predicted SSD delay (with increasing luminosity predicting larger delays) for the given redshift range. This allows for easy interpretation of whether lag dependence on luminosity is steeper or flatter than that of the SSD paradigm. The inherent correlation between luminosity and redshift remains encoded as predicted lags migrate to larger values with decreasing rest-frame wavelength. The dynamic range of continuum luminosity within each $1/(1+z)$ bin sits between 1--1.5 dex as measured by the difference in median $\lambda L_{3000}$ between the most and least luminous bins. 

\defcitealias{2017_rakic_intrinsic}{Rakić et al. 2017}

From Figure \ref{fig:main_results_lag} we qualitatively observe that the measured delays scale as expected with luminosity in the SSD paradigm, and that when controlling for redshift, the previously reported anti-correlation is not observed. As most constituent light curves have approximately the same length, and all objects within a panel have similar redshifts, direct comparisons between lags are valid as we have controlled for the rest-frame wavelengths and timescales probed. These results disfavour the CHAR model for explaining the larger-than-expected lags as, for a fixed observing duration, it predicts delay times should decrease towards the SSD expectation as luminosity increases \citep{2020_sun_char,2021_li_faint,2024_chen_corona-heated}. Previous association of the BLR model with the apparent luminosity-$\tau_{\rm obs}/\tau_{\rm SSD}$ anti-correlation is predicated on a Baldwin effect for nebular emission \citep{2021_li_faint,2022_guo_active}. While broad Balmer lines do not show this behaviour across the general AGN population \citepalias{2017_rakic_intrinsic,2020_rakshit_spectral}, we do detect a weak anti-correlation in our sample for quasars that have valid H\,$\beta$ measurements; however, the magnitude of this effect is small and on the order of the intrinsic scatter in H\,$\beta$ EW. It remains unclear how the expected diffuse contamination changes with luminosity, with the contribution to delays potentially growing with increasing (X-ray) luminosity \citep{2022_vincentelli_luminosity}.

Our observations comparing the stacked ICCF and \textsc{javelin} methods broadly reflect those made in the literature for single object applications. Delay estimates agree under the quoted uncertainties in most bins, with \textsc{javelin} often displaying smaller uncertainties \citep[e.g.][]{2019_edelson_first,2020_yu_reverberation}. We present only \textsc{JAVELIN} measurements for the remainder of this work, preferring this algorithm for its smaller scatter around the a priori assumption delays should follow a power-law scaling and the lower frequency of multi-modal stacked distributions (see Appendix \ref{sec:delay_props} for more discussion).

\begin{figure}
    \centering
    \includegraphics[width=\columnwidth]{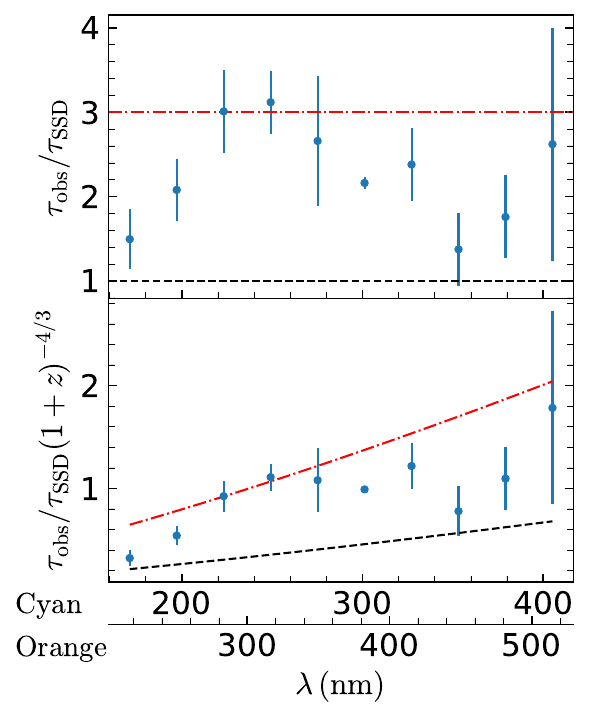}
    \caption{The mean (and RMS) observed-to-predicted ratio in each panel of Fig.~\ref{fig:main_results_lag} against each filter's effective wavelength at the $1/(1+z)$ midpoints. The global wavelength dependence is re-encoded with a $(1+z)^{-4/3}$ term in the bottom panel. The black dashed and red dash-dotted lines again represent one and three times the SSD prediction.}
    \label{fig:lag_amp_ratios}
\end{figure}

We dive further to examine an apparent amplitude trend as we move through our redshift bins. Figure \ref{fig:lag_amp_ratios} showcases the mean delay amplitude relative to the SSD prediction in each redshift bin. With no strong visual evidence that luminosity-scaling deviates from the SSD prediction, we choose this null hypothesis to normalise the delay amplitudes to compare between redshifts and remove the inherent luminosity correlation. The displayed uncertainties in the mean ratios are given by the scatter around the expected relation as there is clearly systematic uncertainties beyond quoted uncertainties (with most panels possessing significant $\chi^2$ values when compared to the mean ratio under the quoted uncertainties). Such systematic uncertainty is likely caused by not controlling for all confounding parameters in our binning. The resulting mean delay amplitudes show a significant, non-monotonic dependence on redshift. Such behaviour is likely a wavelength effect rather than a resonant timescale. This disfavours the \citetalias{2021_kammoun_uv-optical} model as the sole and primary explanation for unexpectedly large disk sizes as it predicts a monotonic relationship with wavelength. Given that Figure \ref{fig:lag_amp_ratios} provides a differential estimate of the lag spectra, it is difficult to rule out whether the underlying disk sizes are inflated and whether the \citetalias{2021_kammoun_uv-optical} model is still applicable. The BLR model, however, offers a natural way to explain both the overall size discrepancy and the non-monotonic amplitude behaviour with wavelength.

Panel 10 of Figure \ref{fig:main_results_lag} shows an elevated lag normalisation consistent with those seen in intensively studied Seyfert galaxies \citep{2016_fausnaugh_space,2019_edelson_first}, but it is unclear whether direct comparison is appropriate given our much longer light curve lengths \citep{2024_edelson_intensive}. Photoionisation models of the BLR predict that the Paschen continuum should contribute significantly to the variable optical-NIR emission \citep{2019_korista_quantifying,2022_netzer_continuum}, and may thus explain the extended delay signals of panel 10. While there is tentative evidence for a Paschen jump in the lag spectrum of NGC 4593 \citep{2018_cackett_accretion}, there is little evidence for a Paschen jump in AGN SEDs. This can be reconciled if the Paschen jump is counterbalanced by a blending of high-order Paschen lines \citep{2022_guo_paschen}, with diffuse emission from this series predicted to make a significant contribution to the total flux \citep{2021_vincentelli_multiwavelength,2022_guo_paschen}.

Photometric contamination from diffuse emission can explain the convex lag amplitude function in $\lambda$ (Figure \ref{fig:lag_amp_ratios}), with a local minima where the ATLAS filters straddle the Balmer jump. The observed delay is minimised when the Balmer continuum contributes significantly to the leading cyan filter and not the orange. Such excess contributions around the Balmer jump have been seen in local type-1 Seyferts through an excess lag signal in the $U$-band \citep{2016_fausnaugh_space,2018_cackett_accretion,2019_edelson_first}, although the degree of this effect is not universal \citep{2021_kammoun_modelling,2023_kara_uv-optical}. Given its presence in our large population study, this suggests contamination from the diffuse BLR emission is widespread. We do not test for delay behaviour as a function of Balmer continuum strength directly as the \citetalias{2020_rakshit_spectral} spectral fitting does not take into account Balmer edge shifts from higher electron number densities and Doppler broadening which can be required in order to accurately measure Balmer continuum contributions \citep{2012_jin_combined}.

The decrease in delay amplitude shortwards of $300\,{\rm nm}$ can be explained by the decreasing contribution to total flux from the diffuse Balmer continuum relative to the disk continuum emission. Ly\,$\alpha$ emission is not expected to contribute substantially to panel 1 (Figure \ref{fig:main_results_lag}) delays as only the higher redshift objects in this bin will probe Ly\,$\alpha$, and the effect will be washed out due to the broad nature of our photometric filters. Detailed photoionisation calculations are beyond the scope of this paper but are necessary for verifying the interpretation given here. These findings are consistent with \citet{2025_pozo_nunez_accretion} where similar rest-frame wavelength delays (to panel 1) are consistent with the SSD model in a similarly bright and massive quasar ($L_{\rm bol}=8.27\times 10^{47}{\rm erg\,s^{-1}}$ and $M_{\rm BH} = 8.9\times10^8M_\odot$) using carefully chosen medium band filters that avoid BLR emission.

\begin{figure}
    \centering
    \includegraphics[width=\columnwidth]{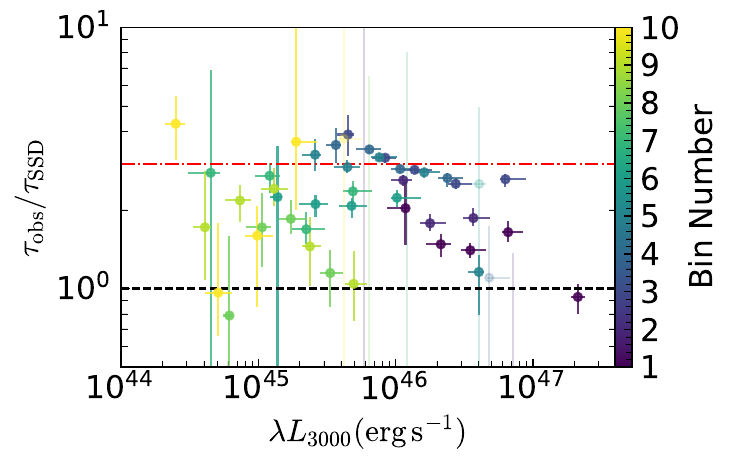 }
    \caption{Log ratio of measured-to-expected lags in each bin against continuum luminosity. Points are colour and opacity coded by their $1/(1+z)$ bin and multi-peaked status respectively, with one and three times the SSD prediction shown as before.}
    \label{fig:lag_ratio_v_luminosity}
\end{figure}

The complex delay amplitude dependence on wavelength also offers a way to reconcile the lack of luminosity-$\tau_{\rm obs}/\tau_{\rm SSD}$ anti-correlation seen here with previous studies. Figure \ref{fig:lag_ratio_v_luminosity} displays the ratio of observed-to-expected delays in the traditional comparison with luminosity. We see an apparent anti-correlation with luminosity when we split before and after panel 6. This anti-correlation is significant for the higher redshift bins, while the lack of significant correlation detected in lower redshift bins is perhaps unsurprising given the lower number statistics and higher uncertainties. This two population behaviour is more evident in our analysis as the close separation of our filters makes us particularly sensitive to deviations from the generally observed $\tau\propto\lambda^{4/3}$ trend \citep{1998_collier_steps,2016_fausnaugh_space}. Previous survey studies that do not explore as high redshifts, and either use more widely separated filters \citep{2019_homayouni_sloan,2024_sharp_sloan} or normalise over several passbands \citep{2020_yu_quasar,2022_guo_active}, are more strongly dominated by the overall wavelength trend and are less sensitive to local deviations, plausibly smoothing out the dependence we see here into a general luminosity anti-correlation.

\subsection{Eddington Ratio Dependence}\label{sec:eddington_lags}

The \citet{1973_shakura_black} disk is the prevailing paradigm for observers modelling accretion in AGN. The model is, in part, predicated on the assumptions of a thin disk ($H\ll R$) that is radiatively efficient (all viscous heat can be dissipated radiatively from the disk surface) and where gas follows Keplerian orbits (there is negligible interaction between neighbouring annuli). For large Eddington ratios ($L/L_{\rm Edd} \geq 0.3$, the requirement to be geometrically thin is violated \citep{1989_laor_massive}. In such highly accreting disks (dubbed slim disks), the local scale height is appreciable (but still $H\lesssim R$) and advective cooling along the equatorial plane becomes significant \citep{1988_abramowicz_slim,2013_abramowicz_foundations}.  

We aim to test for differences in accretion disk structure with Eddington ratio using inter-band time delays. To do this, we limit our subsample to those objects that have high quality (${\rm QUALITY\_MBH=0}$) black hole measurements using either \ion{Mg}{ii} or H$\beta$ lines. We preference \ion{Mg}{ii} derived masses in our high redshift bins as \ion{C}{iv} derived masses can be overestimated, particularly in quasars with high \ion{C}{iv} blueshift \citep{2016_coatman_civ}. This choice also ensures that derived Eddington ratios are comparable across a wider range of bins. We transition to using H$\beta$-derived masses exclusively when $z\leq 0.72$ (panels 6--10) as more quasars have valid black hole measurements in this region (compared to \ion{Mg}{ii}). To separate our subsamples we chose a cut-off of $L/L_{\rm Edd} =0.2$ as significant changes in the ionising continuum behaviour are seen above this value \citep{2023_temple_testing}.

\begin{figure*}
    \centering
    \includegraphics[width=\textwidth]{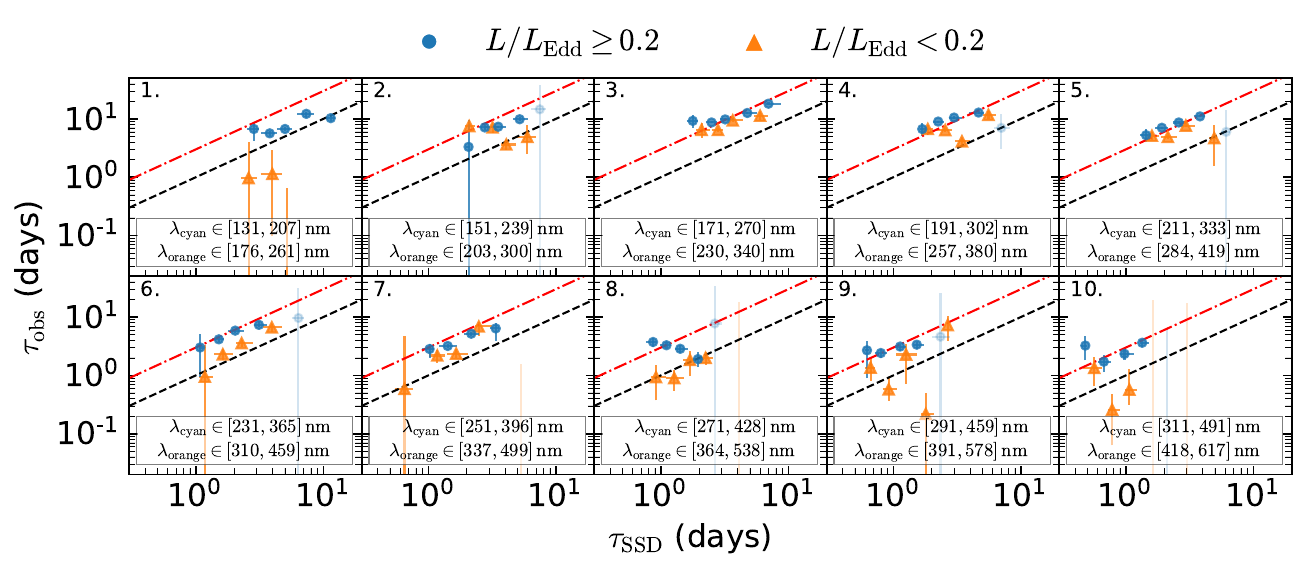}
    \caption{Observed frame stacked \textsc{javelin} delays vs. one (black dashed line) and three (red dash-dotted line) times $\tau_{\rm SSD}$. Delays are split between high (blue circle) and low (orange triangles) Eddington ratio, with low opacity points reflecting multi-modal posterior distributions (Section \ref{sec:multi-peak}). The stated wavelength ranges are the filter extents at each panel's $1/(1+z)$ midpoint.}
    \label{fig:medd_lags}
\end{figure*}

Figure \ref{fig:medd_lags} shows a preference for quasars with higher Eddington ratios to exhibit longer delays. Given that we control for continuum luminosity in our binning, an alternative interpretation is that quasars with larger $M_{\rm BH}$ possess shorter delays. We disfavour this interpretation given the weak predicted dependence on $M_{\rm BH}$ for fixed luminosity (Figure \ref{fig:lag_ratio}), reasoning this behaviour is more likely driven by accretion rate.

Higher Eddington ratio quasars may have longer delays as a result of possessing slim (rather than thin) disks. When the photon diffusion timescale in the vertical direction exceeds the viscous inflow timescale, photons become 'trapped' with their energy being advected into the black hole. The photon trapping radius within which this effect is expected is given by \citep{2006_watarai_new},

\begin{equation}\label{eq:photon_trap_R}
    R_{\rm tr} \approx 102\left(\frac{\dot{m}}{100}\right)R_g
\end{equation}

\noindent where $\dot{m} = \dot{M}/\dot{M}_{\rm crit}=\dot{M}c^2/L_{\rm Edd}=L/\eta L_{\rm Edd}$. In this regime where advective cooling dominates ($R < R_{\rm tr}$), the radiative efficiency decreases relative to a thin disk, with the effective temperature instead following a shallower $T\propto R^{-1/2}$ relation \citep{1999_wang_self-similar,2006_watarai_new}. As a result, less light is emitted from the inner radii and the delay signal is weighted towards larger distances. The degree of this effect is thus dependent on how much of the relevant emitting region falls within the photon trapping radius. This picture is somewhat complicated by slim disks having sufficiently fast radial velocities to allow optically thick thermal emission within the ISCO \citep{1988_abramowicz_slim}, which can emit beyond the peak temperature of the equivalent SSD description for large accretion rates. However, this interpretation is likely still valid as, due to the small spatial size of the region, the overall SED is still redder than that of the thin disk \citep{2000_mineshige_slim}, requiring some net reduction in flux in the inner region that produces the UV continuum.

To reason whether the reduced radiative efficiency of slim disks can explain the lag-Eddington ratio dependence we see, we first estimate the photon trapping radius using Eq. \ref{eq:photon_trap_R}. To convert our previously estimated Eddington ratio to the relevant $\dot{m}$ parameter, we use $\eta=1/25$ following \citet{2000_mineshige_slim} for $L/L_{\rm Edd} < 2$. We then compare the estimated $R_{\rm tr}$ to the responsivity-weighted radius of the orange passband $R_o$ ($c$ times the orange transfer function centroid), finding only 42.4\% of the sample have $R_{\rm tr}/R_o\geq 0.1$, and of those quasars, 89.8\% are in bins 1--3. It is difficult to quantify what effect on the expected delay this ratio will translate to without modelling the temperature profiles explicitly as analytic prescriptions often fail to describe the innermost regions when $\dot{m} < 100$ \citep{2006_watarai_new}. Irrespective, we see that significant fractions of the expected emitting region are expected to be affected by photon trapping at larger redshifts where the ATLAS filters probe the inner radii. As we see examples of larger lags in higher Eddington ratio quasars across the entire redshift range, photon trapping is unlikely to be the sole driver of the effect seen.

It is unclear whether an alternative explanation lies in the relation between disk scale height ($H_{\rm D}$) and Eddington ratio. As Eddington ratio increases, so does the predicted scale height of the disk \citep{1988_abramowicz_slim,1989_laor_massive} but with a concave radial profile ($\partial^2 H_{\rm D}/\partial R^2 \leq 0$). Only in the case of a convex disk scale height profile ($\partial^2 H_{\rm D}/\partial R^2 \geq 0$) do we expect delay times to grow \citep{2023_starkey_rimmed} as the incident angle of X-rays onto the disk becomes stronger at larger radii (as opposed to the concave case where the disk is self-shielding and has the steepest incidence angle in the inner regions). Clearly, more complex modelling is needed to rule out whether Eddington ratio dependent delays can be explained by structural changes in the accretion disk over other potential confounding effects.

Given the large degree of diffuse BLR contamination in our continuum delays, the observed Eddington ratio dependence may only indirectly trace accretion disk changes and instead be primarily driven by changes in BLR structure. Eddington ratio has been identified as the primary driver of spectroscopic diversity in AGN as categorised by the four-dimensional eigenvector 1 (4DE1) space \citep{1992_boroson_emission-line,2000_sulentic_eigenvector,2007_sulentic_eigenvector,2010_marziani_broad-line}. This highlights a key dependence of BLR structure on $L/L_{\rm Edd}$ \citep[possibly through BLR column density;][]{2011_dong_controls}. Detailed theoretical work modelling BLR structure is beyond the scope of this paper but is necessary to help disentangle how much of our measured inter-band delays belong to the BLR and how much residual accretion disk signals will depend on Eddington ratio.

\subsection{Iron Contamination}\label{sec:iron_results}

Motivated by presence of broad line emission in our continuum delays (Section \ref{sec:main_results}), we explore whether the relative strength of iron emission in our quasars influences the recovered inter-band delays. As reverberation mapping lags having been successfully measured for the optical \ion{Fe}{ii} complex \citep{2013_barth_lick}, some fraction of iron emission must respond to variations in the ionising continuum and contaminate our inter-band continuum delays. As one of the most distant BLR regions \citep{2013_barth_lick,2014_chelouche_tentative,2016_marinello_emission}, \ion{Fe}{ii} emission may have an appreciable effect on our measured delays even with smaller EWs (<200\AA). 

\begin{figure*}
    \centering
    \includegraphics[width=\textwidth]{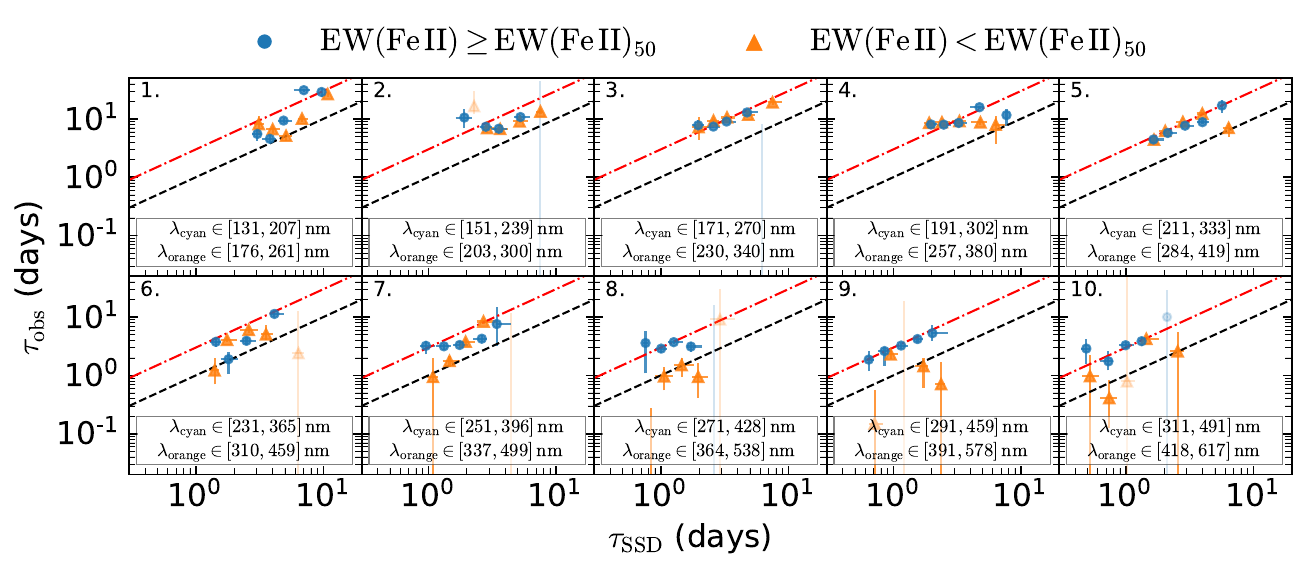}
    \caption{Similar to Figure \ref{fig:medd_lags}, but now split by \ion{Fe}{ii} EW.}
    \label{fig:fe_lags}
\end{figure*}

We measure lags using the \ion{Fe}{ii} EW divided subsets constructed in Section \ref{sec:fe_sample}, finding longer delays in quasars with stronger optical \ion{Fe}{ii} emission and no discernable difference with UV \ion{Fe}{ii} (Figure \ref{fig:fe_lags}). The FWHM of the UV and optical iron complexes are comparable in magnitude and correlated across the AGN population \citep{2008_hu_systematic,2015_kovacevic_connections,2019_le_comparison}, suggesting these two complexes originate from similar regions within the BLR. However, the emission behaviour between the two regimes differ, with the UV iron complex emitting more asymmetrically (beamed towards the centre) due to the higher optical depths and optical iron emission being more isotropic \citep{2009_ferland_implications}, reconciling previous issues with the UV/optical emission ratio \citep{2004_baldwin_origin}. The two regimes are also expected to behave differently at the high columns densities required to explain the systematic redshifts of \ion{Fe}{ii} \citep{2008_hu_systematic,2019_le_comparison}. At these high densities, optical \ion{Fe}{ii} emission (relative to H$\,\beta$) is expected to increase with increasing column density \citep[which is the theorised connector between $L/L_{\rm Edd}$ and BLR spectral properties][]{2011_dong_controls}, whereas UV \ion{Fe}{ii} emission (relative to H$\,\beta$) saturates \citep{2009_ferland_implications}. Reflecting previous literature \citep{2011_dong_controls}, we also see a moderate correlation between optical \ion{Fe}{ii} EW and $L/L_{\rm Edd}$ and weak anti-correlation between UV \ion{Fe}{ii} and $L/L_{\rm Edd}$. As the EWs between the UV and optical complexes are of similar in magnitude and show no redshift evolution (Figure \ref{fig:fe_redshift_evo}), a delay dependence on optical \ion{Fe}{ii} and not UV is likely due to confounding factors from the systemic differences in their physics rather than a differing degree of continuum band contamination to our delays. More theoretical modelling is needed to rule out whether the effect seen here may be due to optical \ion{Fe}{ii} reprocessing.

\begin{figure*}
    \centering
    \includegraphics[width=\textwidth]{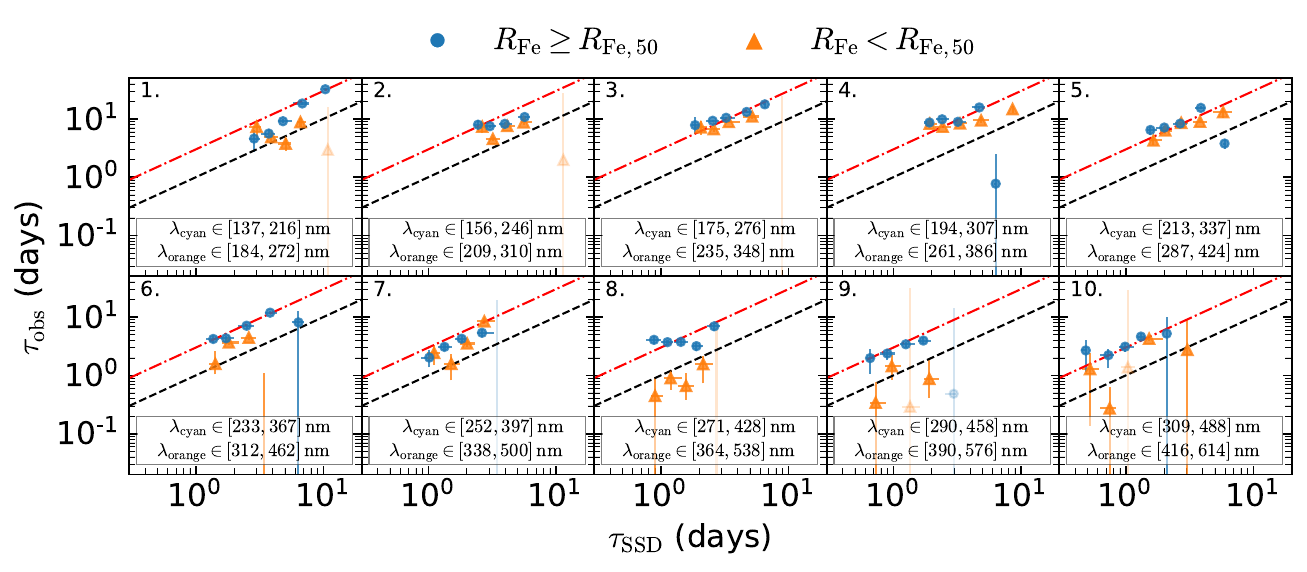}
    \caption{Similar to Figure \ref{fig:medd_lags}, but now split by the ratio of \ion{Fe}{ii} EW to \ion{Mg}{ii} (panels 1--6)/H$\,\beta$ (panels 7--10) EW.}
    \label{fig:metal_lags}
\end{figure*}

\begin{figure}
    \centering
    \includegraphics[width=\linewidth]{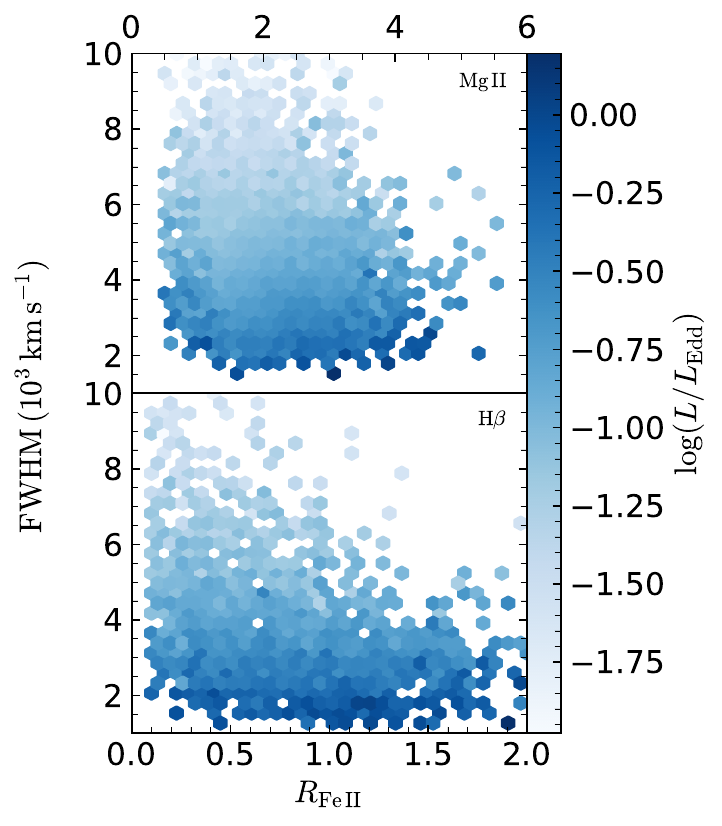}
    \caption{Distribution of quasars in the plane of low-ionisation line FWHM and iron ratio. The distribution is coloured by the median $\log(L/L_{\rm Edd})$.}
    \label{fig:FeII_medd_plane}
\end{figure}

We aim to examine a delay dependence on BLR structure more directly by using $R_{\rm Fe}$, a principal variable in classifying AGN spectroscopic diversity \citep{2000_sulentic_eigenvector,2010_marziani_broad-line}. We measure delays for the $R_{\rm Fe}$ divided subsets established in Section \ref{sec:fe_sample}. As expected (due to the inherent correlation with $L/L_{\rm Edd}$), we find longer lags in quasars with higher \ion{Fe}{ii} ratios (in both the optical and UV regimes; Figure \ref{fig:metal_lags}). Figure \ref{fig:FeII_medd_plane} shows the distribution of Eddington ratios in our quasars for the optical plane of the 4DE1 space and its UV analogue. Broadly following theoretical predictions for the optical diagnostics \citep{2002_zamanov_searching}, we see our quasars are in a regime where Eddington ratio is more strongly separated by H$\,\beta$ FWHM than $R_{\rm Fe}$. We still do see appreciable Eddington ratio separation by subdividing at the median optical $R_{\rm Fe}$ value, confirm that the distributions and mean values of $\log(L/L_{\rm Edd})$ are significantly different with a Kolmogorov-Smirnov test and two sample t-test respectively. We note that this separation in Eddington ratio is also present with UV $R_{\rm Fe}$, but to a much lesser degree, perhaps owing to the weaker relationship between the UV \ion{Fe}{ii} emission and Eddington ratio.

Despite the much weaker relationship with Eddington ratio, the UV $R_{\rm Fe}$-divided subset achieves a similar magnitude effect on inter-band delays to dividing directly by Eddington ratio (comparing panels 1--6 in Figures \ref{fig:medd_lags} and \ref{fig:metal_lags}). As the ratio of UV \ion{Fe}{ii} EW to \ion{Mg}{ii} EW may serve as a proxy for chemical enrichment \citep{2011_de_rosa_evidence}, this may signal a metallicity dependence in our delays \citep[although this ratio is dependent on excitation conditions in the BLR and not solely on Fe abundance;][]{2004_baldwin_origin}. Disks with high iron content may have inherently different disk structure \citep{2016_jiang_iron} with more powerful intrinsic disk variability \citep{2020_jiang_opacity}. If such a metallicity dependence is present, it runs counter to the previously seen anti-correlation of continuum delays with \ion{Fe}{ii}/\ion{Mg}{ii} EW ratio \citep{2017_jiang_detecting}. We however have a much larger sample size to simultaneously control for luminosity and wavelength effects. Unfortunately, we lack the sensitivity to further divide our sample to separate out the potential UV $R_{\rm Fe}$ effect from the $L/L_{\rm Edd}$ effect and any other BLR structural dependencies.

\subsection{High Outflow Lags}\label{sec:civ_results}

The physical conditions in AGN can give rise to outflowing winds as material is ejected from the accretion disk. Such a model is generally used to explain the blueshifted, \ion{C}{iv} absorption \citep{1995_murray_winds} and emission \citep{2004_leighly_hubble,2011_richards_unification} features seen in the ultraviolet spectra of AGN. Evidence for \ion{C}{iv} outflows is often more strongly present in high mass, high Eddington ratio AGN \citep{2020_rankine_bal,2023_temple_testing} where the ionising continuum is softer (as indicated by decreased \ion{He}{ii} EW) and radiation line-driving mechanisms are likely dominant \citep{2019_giustini_global}.

The presence of disk winds may help explain the inter-band delay behaviour seen here and in the literature. The ejection of material from the disk creates a radially dependent mass-accretion rate, flattening the temperature profile and requiring hotter overall normalisation in the outer disk to recover observed luminosities. In this picture, a greater fraction of the variable and static emission comes from more distant radii than expected in the SSD model, explaining the inflated size measurements \citep{2019_sun_winds, 2019_li_reconciling}. Alternatively, this spectral reddening could shorten delays, with a weaker ionising continuum (for fixed UV-optical luminosity) lessening the contribution from variable BLR emission to continuum band delays \citep{2025_netzer_disc}. 

The effects of winds on continuum-band delays are generally discussed in the context of quasars with broad absorption lines (BAL quasars) in the UV or high column density obscuration in the X-ray. Both properties are variable on the timescales of days to years, with the two often varying in tandem \citep{2016_ebrero_discovery,2022_mehdipour_transformation,2023_partington_storm}. These obscuring winds have been linked to longer continuum \citep{2024_lewin_storm,2025_lewin_accretion} and UV BLR delays \citep{2024_homayouni_storm}. This correlation is interpreted as the ejected material reprocessing the incident X-ray/disk emission \citep{2025_lewin_accretion} which has the potential to explain the increased delay signal see in AGN \citep{2023_hagen_modelling,2024_hagen_variability}. 

Given the lack of available time-coincident observations of the X-ray obscuring and UV absorbing wind for our large sample, we test for disk wind effects in our lags via a proxy: strong \ion{C}{iv} emission line outflows. Reconstructed BAL quasar \ion{C}{iv} emission line profiles are co-located with their non-BAL counterparts in \ion{C}{iv} EW-blueshift space, with a corresponding non-BAL quasars available with equivalent properties (luminosity, black hole mass, Eddington ratio, \ion{He}{ii} EW) for each BAL quasar \citep{2020_rankine_bal}. This suggests that these two classes of quasars are not distinct but come from the same underlying population. Observing a broad absorption trough may thus be a phase or time dependent property with several AGN seen transitioning between BAL and non-BAL states \citep{2012_filiz_ak_broad,2018_rogerson_emergence,2019_sameer_x-ray}. By splitting our sample by \ion{C}{iv} emission line blueshift, we hope to preferentially select a population of quasars with the largest potential effect should they enter a BAL state, as the strongest absorption troughs are seen in quasars with more blueshifted \ion{C}{iv} emission \citep{2020_rankine_bal, 2022_rodriguez_hidalgo_connection}. 

For our subsample of quasars with good quality \ion{C}{iv} blueshift measurements, we split our sample according to Section \ref{sec:civ_sample}. This maintains reasonable sample sizes in panels 1--3 of our binning scheme. We redefine our five uniform $\log(\lambda L_{3000})$ bins to reflect the changes in luminosity range. Without redistributing objects, we do not have sufficient sample counts in our lowest and highest luminosity bins. Having only three bins per rest-frame panel inhibits our ability to assess if trends are real or spurious under the uncertainties. High \ion{C}{iv} blueshift bins do have notably lower counts across the parameter space.

\begin{figure*}
    \centering
    \includegraphics[width=\textwidth]{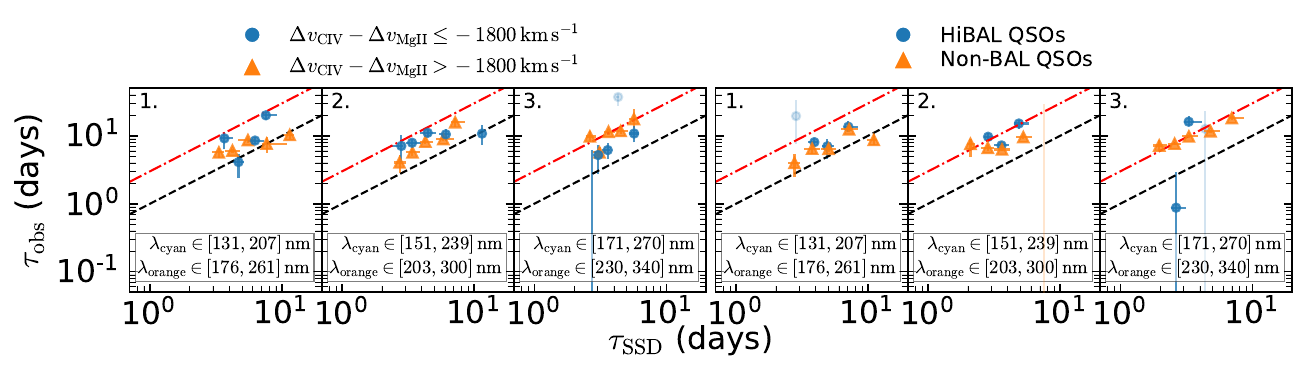}
    \caption{Similar to Figure \ref{fig:medd_lags}, now splitting by \ion{C}{iv} emission line blueshift (left) and BAL status (right).}
    \label{fig:wind_lags}
\end{figure*}

Figure \ref{fig:wind_lags} (left) compares delays measured in quasars with and without strong \ion{C}{iv} emission line outflows. Overall, delays are consistent with the better-constrained main sample, although there is larger scatter in the more blueshifted \ion{C}{iv} bins likely owing to the smaller number statistics. A slight preference for longer lags in quasars with more blueshifted \ion{C}{iv} emission is seen in panels 1 and 2, with an inversion of this behaviour in panel 3. Such rapid changes in behaviour at adjacent redshifts imply a localised wavelength affect, perhaps stemming from the direct contamination of the cyan filter by the \ion{C}{iv} line. As \ion{C}{iv} blueshift is anti-correlated with equivalent width, it is difficult to disentangle the effects of the two line properties for contaminated redshift bins. If the default behaviour implies longer lags in quasars with less blueshifted (i.e. larger EW) \ion{C}{iv} emission (i.e. panel 3), then the shorter delays observed in panels 1 and 2 may stem from the greater contribution from the more distant BLR to the leading cyan band. This conclusion is difficult to verify as there is only one uncontaminated redshift bin with corresponding \ion{C}{iv} information available.

Physical explanations for longer lags in our low blueshift populations may lie in the link between \ion{C}{iv} morphology and the SED of the EUV continuum \citep{2023_temple_testing}. \ion{C}{iv} blueshift is anti-correlated with the \ion{He}{ii} EW \citep{2015_baskin_origins,2020_rankine_bal,2023_temple_testing} which, as a recombination line, serves as a proxy for the number of ionising photons above 54eV. Thus, higher \ion{C}{iv} may signal a softer/less luminous ionising continuum, which would result in a smaller $L_{\rm diffuse}/L_{\rm disk}$ and measured delay \citep{2025_netzer_disc}. A delay dependence on other ionising continuum proxies ($L[$\ion{O}{iii}$]/L({\rm H}\,\beta)$) has not been observed however \citep{2024_sharp_sloan}. Alternatively, \ion{He}{ii} EW may probe the BLR covering fraction seen by the EUV continuum \citep[although the former interpretation is favoured; see discussion in][]{2013_baskin_average}. Quasars with less blueshifted \ion{C}{iv} emission (weaker \ion{He}{ii} EW) would then have larger BLR covering fractions and thus longer delays \citep{2022_netzer_continuum}. Interestingly, we confirm with a non-equal variance two-sample t-test that the high \ion{C}{iv} blueshift objects have, on average, larger Eddington ratios. Again, with only one band not directly probing \ion{C}{iv} emission it is difficult to ascertain whether this is a universal trend acting in opposition to the previously seen Eddington ratio dependence (Section \ref{sec:eddington_lags}) and the \ion{C}{iv} connection to the 4DE1 space \citep{2000_sulentic_eigenvector,2007_sulentic_eigenvector}. 

We discuss an alternate hypothesis that longer lags are seen in our low \ion{C}{iv} blueshift sample is due to contamination from BAL quasars. While the fraction of quasars with broad absorption troughs increases with \ion{C}{iv} emission line blueshift for fixed EW, this fraction peaks below our chosen cut-off blueshift \citep[see Figure 10 of][]{2020_rankine_bal}. While quasars displaying broad absorption troughs in \ion{C}{iv} where removed from this subsample, as this phenomenon is time-varying, our \ion{C}{iv} separated sample may achieve a greater current fraction of BAL quasars in weakly blueshifted \ion{C}{iv} bins. This may explain our previous interpretation of longer delays in these bins with previous evidence disk wind obscuration increases delay times \citep{2024_lewin_storm,2025_lewin_accretion}.

We attempt to probe delays in BAL quasar directly from their historical SDSS flags. HiBALs, that show absorption troughs for high-ionisation lines such as \ion{C}{iv}, make up only 7\% of our sample in the three highest redshift bins (with no information available at lower redshifts). We compare lags derived from only these historic BALs to a non-BAL sample (removing LoBALs, HiBALs, and those whose status was unmeasurable) in Figure \ref{fig:wind_lags} (right). Results are broadly consistent with HiBAL quasars exhibiting longer delays at the available wavelengths, although these delays are less well constrained with such low sample counts (some equivalent HiBAL bins are empty). Further dispersion is expected as we use potentially outdated information about the time-variant BAL states. As our \ion{C}{iv} information is obtained from observed-frame optical spectra, we are unable to investigate whether this behaviour extends to longer wavelengths as has been seen in other studies \citep{2024_lewin_storm,2025_lewin_accretion}, highlighting the challenges of survey-based investigations into disk wind properties. Clearly, future continuum RM campaigns with coincident disk wind measurements and more precise/wider wavelength coverage photometry are needed to investigate the role of disk winds in inter-band delay measurements.

\subsection{Colouring Lags by Residual Colour}\label{sec:colour_lags}

While the standard model of \citet{1973_shakura_black} accretion disks broadly explains the shape and strength of AGN UV-optical continua, it fails to capture spectral behaviour in detail. The NUV and FUV continuum slopes of AGN are often redder than predicted \citep{2007_bonning_accretion,2007_davis_continuum} and appear bounded below the expected $\alpha=1/3$ ($f_\nu\propto\nu^\alpha$) slope \citep{2016_xie_luminosity}. Potential reasons for the unexpectedly red continua include advection cooling in the disk \citep{2014_netzer_bolometric,2019_kubota_modelling}, disk-wind outflows \citep{2012_slone_effects,2014_laor_winds}, and dust reddening \citep{2003_richards_red,2015_capellupo_active,2016_baron_evidence,2016_xie_luminosity}, all of which may induce longer lags. As such, we search for a colour dependence in continuum delays here.

\begin{figure*}
    \centering
    \includegraphics[width=\textwidth]{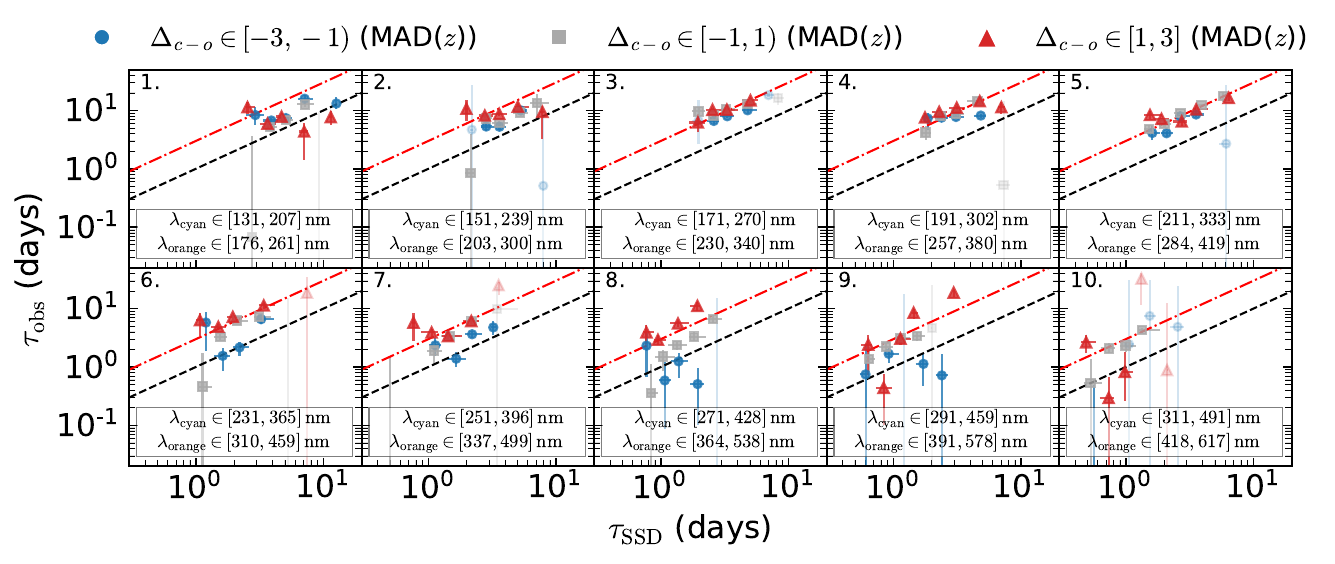}
    \caption{Similar to Figure \ref{fig:medd_lags}, now split by quasar residual colour into bluer, average, and redder quasars, respectively.}
    \label{fig:colour_lags}
\end{figure*}

In Section \ref{sec:remove_dust}, we established a broad distribution of colour in our sample relative to a redshift dependent median. After removing the red-tail, this distribution is approximately Gaussian. We divide this residual colour distribution into three uniform bins with the resulting lags are shown in Figure \ref{fig:colour_lags}. We see delays are often ordered by their residual colour, with redder quasars possessing longer lags. The magnitude and coherence of this observed effect is not uniform across our parameter space. The largest scatter is seen in the lowest redshift bins where we have the smaller number statistics (binning in $1/(1+z)$) and residual host-galaxy contamination is most likely present. It is difficult to disentangle whether this behaviour is separate from that seen in Section \ref{sec:eddington_lags} as several redshift bins have significant Spearman rank correlations between Eddington ratio and residual colour (higher Eddington ratio quasars are redder). With the larger scatter observed here in the colour-separated delays, further subdividing this sample by multi-collinear explanatory variables is beyond the limit of our data.

Internal AGN reddening presents a potential explanation for these colour-dependent lags. Redder residual colours would then indicate AGN experience a larger degree of dust extinction and have their luminosity more greatly underestimated \citep[and hence underestimating their delay times;][]{2017_gaskell_case,2023_gaskell_estimating}. As we control for the observed luminosity, this presents as redder quasars having longer delays in their respective panels. The hypothesis of dust reddening driving the observed delayed behaviour would be strongly supported if the red quasars excluded in Section \ref{sec:remove_dust} exhibited the longest lags. Quasars belonging to the red-tail of the residual colour distribution are thought to be strong evidence of dust extinction in AGN \citep{2003_richards_red}. As a consequence of this dust reddening, these very red quasars are among the faintest in our sample with many hovering just above the minimum Gaia $R_p$ magnitude ($R_p\leq18$). As Gaia $R_p$ occupies similar wavelengths as the ATLAS orange band, these very red quasars will be even fainter in the cyan band, which is the limiting passband in this analysis. As a result, there is insufficient signal to constrain delays for quasars in the dust-reddened tail using ATLAS data, even with our stacked methodology. The future Legacy Survey of Space and Time \citep[LSST;][]{2019_ivezic_lsst} will offer a fantastic opportunity to study these very red quasars with its superior photometric sensitivity. If dust reddening was responsible for colour dependent delays, then this may open up the exciting prospect of using continuum delays to measure reddening laws in quasars. 

With evidence of diffuse nebular emission contaminating our delays, the observed colour-dependent lags may alternatively be explained by a varying BLR covering factor. Quasars with larger BLR covering factors would possess more strongly emitting BLR regions relative to the accretion disk, lengthening delays \citep{2022_netzer_continuum}. Such spectral signatures would also then be expected in the time-average SED and photometric colours. In this picture, our colour dependence has a more nuanced interpretation (unlike with the dust reddening case which is largely monotonic with wavelength), as our colours are derived at fixed observed-frame wavelength. As the Balmer and Paschen continua are redder than the underlying disk continuum, when both filters probe the same nebular emission, quasars with larger BLR covering factors would appear redder and have longer delays. When filters straddle the Balmer jump, quasars with larger covering factors would have larger nebular emission contributions to the cyan filter, making the time-averaged photometric colours appear more blue and shortening the observed inter-band delay. As we only have two filters, the differential delay operator that provides evidence for this non-monotonic wavelength behaviour (Figure \ref{fig:lag_ratio}) does not rule out larger delay signatures in an absolute sense. For example, redder quasars in panel 8 of Figure \ref{fig:colour_lags} are seen to be at least a factor three larger than anticipated in the SSD. This same (spanning the Balmer jump) wavelength range is where the observed-to-predicted delay ratio experiences its minima in our main sample delays, perhaps signalling that the underlying disk lags (with less BLR contamination, i.e. redder residual colours at this redshift) are still larger than previously anticipated. This difference is more easily reconciled without revisions to accretion disk theory in the dust extinction picture which behaves similarly at all wavelengths. 

\section{Summary and Outlook}

We have analysed high-cadence ATLAS light curves to measure some of the most precise population-level inter-band continuum delays for $\sim$9,500 of the most UV-optical luminous quasars in the sky. With our large sample size, we were able to correlate delays over a range of quasar properties and help break degeneracies between several population literature models. We summarise the main findings of this work as follows:

\begin{enumerate}
    \item We test our ability to recover continuum band delays on simulated data and demonstrate that selecting delays based on significance introduces a bias towards longer lags (Figure \ref{fig:sim_comparison}). When we combine posterior estimates without significance cuts, no bias is introduced, but with our data, the delay uncertainties are very large. Only by stacking delay information at the inference level, rather than combining individual posteriors, do we greatly improve our ability to measure the underlying signal. Some benefit is seen on the simulated dataset to analysing only the more well correlated light curves, but we show that the inter-band correlation is dependent on quasar colour (relative to a redshift dependent median; Figure \ref{fig:col_res_hexbin}). As such, cutting by correlation biases our sample towards bluer quasars.
    
    \item Using the stacked methodology, we compile lags across a range of redshifts ($z\sim0.3-2.5$) and luminosities ($\lambda L_{3000}\sim10^{45}-10^{47}{\rm erg\,s}^{-1}$). When we compare quasars within a given redshift bin, we do not see an anti-correlation between the disk size discrepancy and luminosity (Figure \ref{fig:main_results_lag}). These findings disfavour the CHAR model to explain our delays which predicts the aforementioned anti-correlation when controlling for light curve duration and wavelength as done here. We further find a non-monotonic trend of delay amplitude with rest-frame wavelength, disfavouring the \citetalias{2021_kammoun_uv-optical} model as the primary source of elevate disk sizes as it predicts a monotonic change. We find our results are naturally explained by diffuse contamination from the BLR, suggesting it is commonplace in the wider AGN population.

    \item After separating our sample by Eddington ratio, we observed that quasars with higher Eddington ratios often prefer longer lags (Figure \ref{fig:medd_lags}). This behaviour is observed across our redshift range suggesting it is not solely driven by the advection cooling of slim disks, which is expected to contribute significantly only at higher redshifts in our sample. The increasing delay magnitude is also unlikely to be from changes in disk scale height with increasing Eddington ratio as the scale height profiles in slim disks are predicted to be concave \citep{1988_abramowicz_slim,1989_laor_massive} and thus produce shorter delays \citep{2023_starkey_rimmed}. Instead, given the strong evidence for diffuse contamination in our delays, the observed Eddington ratio dependence like stems from its connection to BLR structure and the 4DE1 space \citep{2007_sulentic_eigenvector,2010_marziani_broad-line}.

    \item We find evidence that delay magnitude increases with optical \ion{Fe}{ii} EW, but not UV \ion{Fe}{ii} EW. Given the similar equivalent width of the two regimes (and the lack of redshift dependence; Figure \ref{fig:fe_redshift_evo}), we believe this is less likely to be evidence of a direct response by the iron BLR, but instead related to BLR structure as UV \ion{Fe}{ii} is less dependent on $L/L_{\rm Edd}$ than the optical complex \citep{2011_dong_controls}. We further find longer delays in quasars with larger $R_{\rm Fe}=$ \ion{Fe}{ii}/H$\,\beta$ (and the UV analogue \ion{Fe}{ii}/\ion{Mg}{ii}) which is a principal variable in explaining AGN spectroscopic diversity \citep{2000_sulentic_eigenvector,2007_sulentic_eigenvector}. As we see a similar magnitude effect with \ion{Fe}{ii}/\ion{Mg}{ii} as $L/L_{\rm Edd}$, while the two are more weakly related (Figure \ref{fig:FeII_medd_plane}), this may signal a possible metallicity dependence in our lags \citep{2011_de_rosa_evidence,2019_shin_ratio}.

    \item Motivated by recent observations linking disk winds to excess continuum delays \citep{2024_lewin_storm,2025_lewin_accretion}, we test to see if delays change with \ion{C}{iv} emission line blueshift. We see evidence that quasars with smaller \ion{C}{iv} blueshifts have longer lags, with this relationship inverting when the \ion{C}{iv} line enters the cyan band (Figure \ref{fig:wind_lags}, left). However, this relationship is difficult to verify with only one panel not directly probing \ion{C}{iv} emission and considering its inconsistency with the Eddington ratio and 4DE1 interpretation. We additionally examine the effect of broad absorption lines by comparing delays between historically confirmed BAL and non-BAL quasars. We find some evidence of longer delays in BAL quasars, although this relation is more tenuous given the low sample counts and transient nature of BALs. This transient nature may help explain our \ion{C}{iv} results as BAL frequency is highest in the low \ion{C}{iv} blueshift subset of our division of the \ion{C}{iv} EW-blueshift plane \citep{2020_rankine_bal}.

    \item We find a tendency for quasars with redder continua to have longer lags (Figure \ref{fig:colour_lags}). There are several plausible mechanisms that can create redder continua, including: slim disks, disk winds, and dust reddening. We explored the first two mechanisms independently (see points iii and v respectively). We are unable to quantify whether residual colour or Eddington ratio is the driver behind this trend, with both potentially being explained through either dust reddening or a variable BLR covering factor. The hypothesis of dust reddening would be more strongly supported if quasars in the dust-reddened tail displayed the longest lags. However, we are unable to further test this hypothesis as these highly reddened quasars have the lowest SNR.
    
\end{enumerate}

We have shown that lags may depend on several physical parameters and disentangling them will require more precise measurements over a larger parameter space than given here. While we have shown how to extract a greater degree of signal from light curves that individually struggle to constrain lags, testing for all the different dependencies is expensive and non-feasible (only so much independent information can be extracted). The upcoming LSST survey will provide an excellent opportunity to further study inter-band continuum delays and correlate them with several physical parameters simultaneously as its superior photometric precision will allow for measuring delays in thousands of AGN individually \citep{2022_kovacevic_lsst,2023_pozo_nunez_modelling}. The deeper photometry of LSST will also allow us to examine delays in the dust-reddened tail of the AGN colour distribution, further shedding light onto the potential role dust extinction in continuum RM.

\section*{Acknowledgements}

We kindly thank the anonymous referee for their insightful comments and assistance in improving this manuscript. KWS acknowledges funding from the Royal Society. AC acknowledges the support from ANID, Chile, through grants AIM23-0001 (Millennium Science Initiative) and FONDECYT No. 1251692.
This work uses data from the University of Hawaii's ATLAS project, funded through NASA grants NN12AR55G, 80NSSC18K0284, and 80NSSC18K1575, with contributions from the Queen's University Belfast, STScI, the South African Astronomical Observatory, and the Millennium Institute of Astrophysics, Chile.

\section*{Data Availability}

The data underlying the analysis in this paper may be shared upon reasonable request.



\bibliographystyle{mnras}
\bibliography{ATLAS_QSO_RM} 




\appendix

\section{Delay Pipeline Dependency}\label{sec:pipeline_check}

\begin{figure*}
    \centering
    \includegraphics[width=\textwidth]{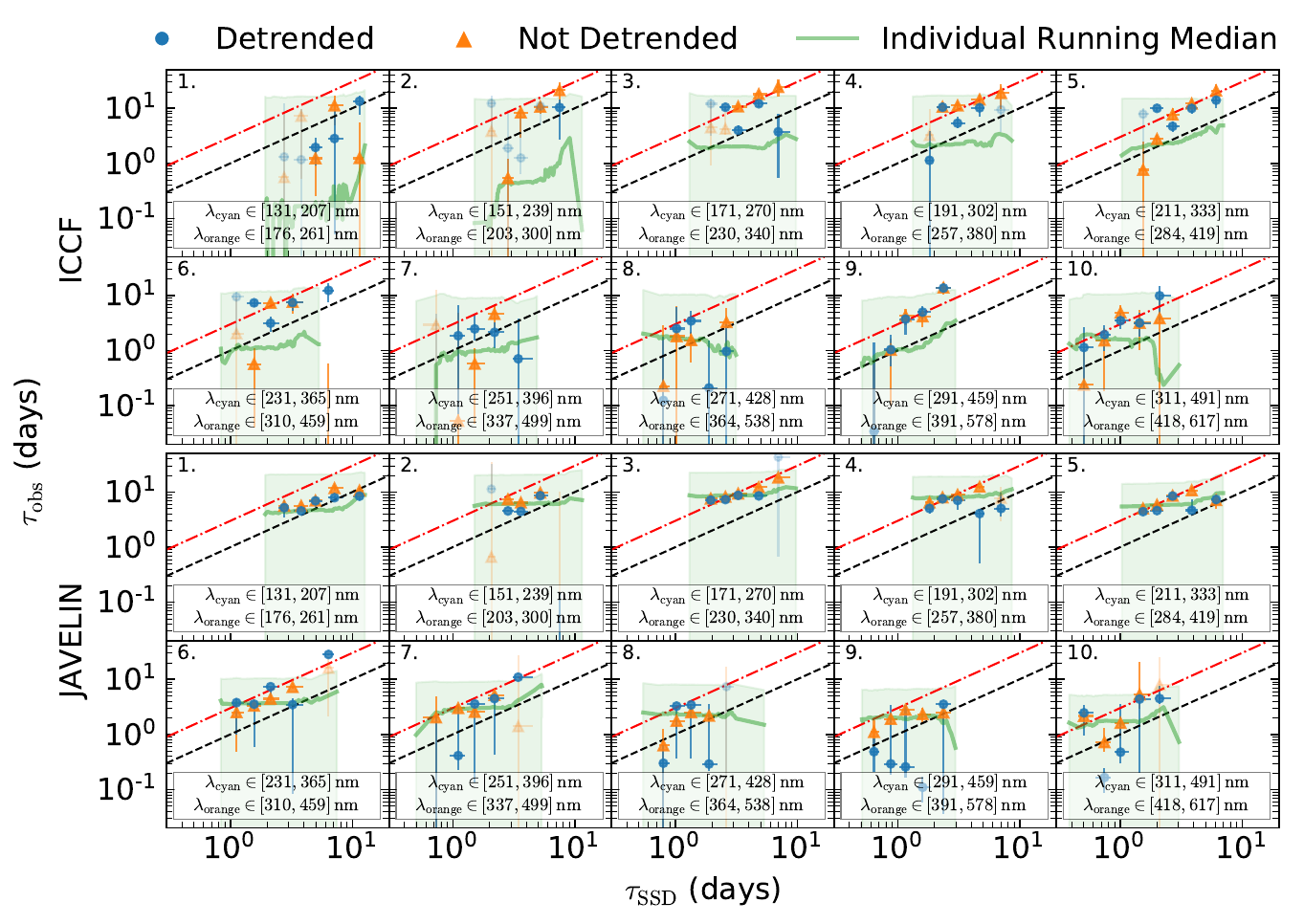}
    \caption{Comparison between the detrended and not detrended stacked delay approaches for ICCF (top) and \textsc{JAVELIN} (bottom) for the main sample. A running median (and 16th and 84th percentiles) derived from applying each algorithm to the individual light curve pairs is also displayed. Delays with multiple peaks in their posterior distributions have reduced opacity.}
    \label{fig:pipeline_comparison}
\end{figure*}

We investigate what influence our analysis pipeline choices have over the measured delays to ensure that we are not introducing any systematic biases through these decisions. We choose the main sample as our representative validation set for the real data. This sample is where we have the greatest number statistics and thus where any potential biases may be best separated from noise. Two possible concerns include whether homogenous detrending of light curves removes genuine physical variability and whether binning without concern for cadence suppresses short-term features. Both, if present, would distort the correlation structure and hence our inter-band delays. As such, we display our fiducial choices alongside two other comparative applications of ICCF and \textsc{JAVELIN} with/without binning and without stacking (Figure \ref{fig:pipeline_comparison}).

Results from ICCF seem to qualitatively prefer the detrended approach in more panels (1,6--10) than not (2--5) based on the a priori expectation that delays follow a power-law scaling with luminosity at fixed redshift. We additionally compare ICCF performance with and without detrending on the simulated dataset and find detrending is necessary for lag recovery, reflecting earlier recommendations in the literature \citep{1999_welsh_reliability}. This detrending helps remove the low frequency components of light curves, which contribute more significantly than high-frequency components to the construction of the CCF \citep{1999_welsh_reliability,2001_vio_limits}, limiting our ability to constrain delays on short timescales. We find detrending important here as a form of local normalisation, which helps reduce spurious correlations in time series with large autocorrelation \citep{2016_dean_dangers}, as is common for systems like AGN where physical limitations of variability impose dependence on previous epochs \citep{2009_kelly_variations}. 

In Figure \ref{fig:pipeline_comparison}, we see \textsc{JAVELIN} performs better on light curves without detrending. Detrending here appears to introduced a systematic bias towards smaller lags (most notably in panels 5--10). Again, this behaviour is reflected in the simulated data. \textsc{javelin} is biased when additional sources of variability are present that are not described by the assumed linear response (one light curve is a shifted, smoothed, and scaled version of the other; Eq. \ref{eq:linearRM}). As we detrend each filter independently with a quadratic fit, we may be artificially introducing additional long-term variability outside of the model assumption, particularly in noisier light curves where the common long term trends may be masked (we see notable decrease in inter-band correlation when detrending; see Figure \ref{fig:rs_comparison}).

Our findings suggest that each algorithm does not optimally perform on the same light curve set: with or without detrending. Ideally, we would compare these methods on the same set of curves as detrending removes some longer term variability that may be associated with reprocessing from the BLR \citep{2022_cackett_frequency,2023_lewin_x-ray} or intrinsic disk variability \citep{2003_uttley_correlated,2009_breedt_long-term,2020_hernandez_santisteban_intensive} that propagates oppositely to X-ray reprocessing lags \citep{2023_yao_negative,2024_neustadt_storm}. We find that this mismatched comparison is necessary to compute reliable lags with each method. We also note that these recommendations are also verified on simulated data for which only the disk reprocessing scenario is present. Given this, and the overall consistency between detrending and not detrending with each algorithm, we do not see strong evidence that choosing one or the other significantly biases the presence of other delay phenomena.

Previously, we discussed (see Section \ref{sec:bin_stack}) how selecting individual inter-band delays based on significance cuts, inter-band correlation, and fractional variability can bias the population distribution of lags. Thus, to compare our stacked lags to the individual object measurements, we compute running median (and 16th and 84th percentile) trends based on the complete, individual lag population. Individual delays are measured using the optimal detrending choice for each algorithm in the stacking scenario. Figure \ref{fig:pipeline_comparison} shows that the joint distribution of individual \textsc{javelin} delays is remarkably similarly to the stacked approach, but with much larger uncertainty. While the distribution of individual ICCF delays broadly matches those of the stacked approach under the uncertainties, the individual estimates do seem to favour shorter delays. It is difficult to verify whether stacking ICCF is smoothing out short timescale structure in the CCF and lengthening delays without also testing a finer time step resolution in the CCF, but given the consistency with the \textsc{javelin} lags, we find it more likely that the individual estimates are biased.

Given our success with inferring joint delays, we test whether delays can be further improved by renormalising light curves to a common scale (by dividing by their maximum value). Renormalisation does not affect ICCF as light curves are renormalised in calculating the Pearson correlation coefficient. For \textsc{JAVELIN}, we observe this renormalisation has an adverse effect on the recovered delays in both the real and simulated data. This effect is more notable in light curves without detrending as we may artificially inflate the importance of light curves that do not show strong long term variability, with the shorter timescale fluctuations being closer to the order of the photometric noise, and hence diluting the signal.

\section{Light Curve Properties and their Influence on Joint Delays}\label{sec:delay_props}

During our analysis, we identified that jointly fit delays were poorly constrained in some bins. These bins are identified by either possessing much larger uncertainties than their peers and/or multiple peaks in the posterior delay distributions. We compile all the lags measured during this work on real data finding approximately $27\%$ of ICCF and $11\%$ of \textsc{JAVELIN} delays possess more than one posterior peak (as defined in Section \ref{sec:multi-peak}). The $<9\%$ of bins with delay semi-amplitude uncertainties ($\Delta\tau_{84-16}/2$) larger than 10 days belong almost exclusively to this multi-peaked population.

We investigate which light properties influence whether measured delays are poorly constrained by fitting a random forest model to predict whether a delay distribution will have multiple peaks using properties such as the number of constituent objects, the level of long and short timescale inter-band correlation, and the median relative uncertainty, fractional variability, and the median and minimum cadence in each passband. The models achieve Area Under the Curve (AUC) scores on our validation set of 0.84 and 0.90 for ICCF and \textsc{JAVELIN} respectively. We compute permutation feature importances for each model, identifying important predictors in long timescale inter-band correlation and median relative photometric error for \textsc{JAVELIN} and short timescale inter-band correlation for ICCF. We see visually that the distributions of these predictors between uni- and multi-modal delay populations differ and confirm with a Kolmogorov-Smirnov test. For \textsc{JAVELIN}, multi-peaked behaviour occurs in regions with high inter-band correlation and low median relative photometric error. While these regions are where we would expect to best constrain lags, the additional important requirement is that the bin has small number of objects ($\lesssim 20$). We compare the composite distribution of individual \textsc{JAVELIN} runs to the joint fit for several offending bins finding that \textsc{JAVELIN} either begins to resolve the individual delay signals or struggles to converge when the individual signals are confidently disparate and few in number. Whether this disparate behaviour spuriously arises from our random sampling of an unfavourable variability episode or some uncontrolled binning parameter is unclear. At this stage, there does not appear to be any systematic issues with \textsc{JAVELIN} that are not explained by poor sample statistics. For ICCF, multi-modal posteriors occur regardless of the number of constituent objects, primarily when short timescale inter-band correlation is low and spurious peaks in the CCF become more common. As such, ICCF does not perform equally well across our parameter space, disfavouring faint and redder object and reaffirming our choice in \textsc{JAVELIN} as the preferred lag algorithm.

\begin{figure}
    \centering
    \includegraphics[width=\linewidth]{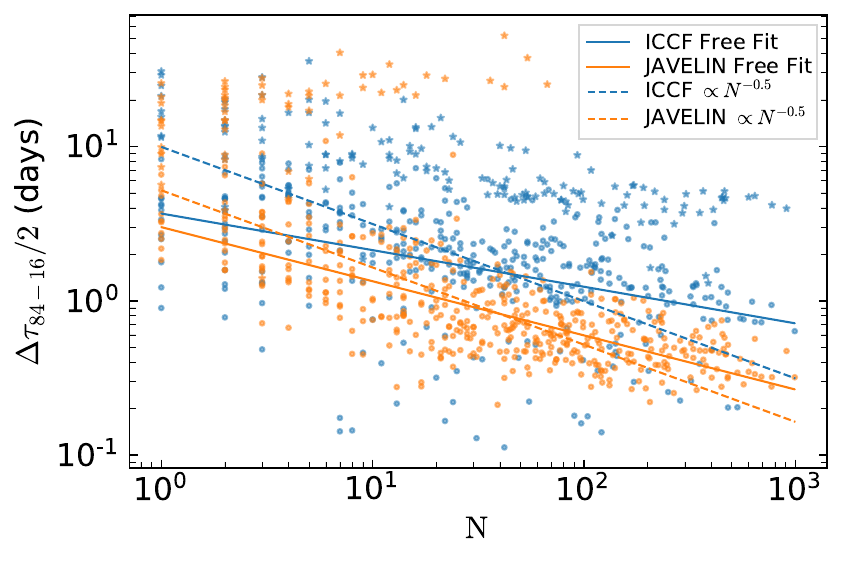}
    \caption{Semi-amplitude (uncertainty) of measured stacked delays against the number of contributing light curves pairs in each bin using both ICCF (blue) and \textsc{JAVELIN} (orange). Power laws with freely varying amplitude and slope are fit to each algorithm (solid lines), along with a fixed $N^{-1/2}$ relation intersecting the median error (dashed lines). Bins where posterior lag distributions showed more than one peak are indicated by stars rather than circles.}
    \label{fig:lag_error_v_N}
\end{figure}

\begin{figure}
    \centering
    \includegraphics[width=\linewidth]{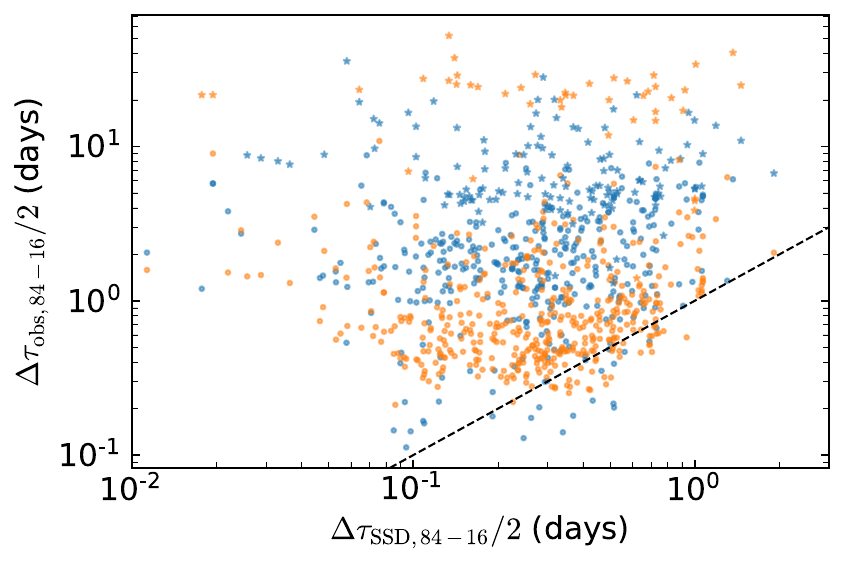}
    \caption{Comparison of the semi-amplitudes measured by the stacked approach to those derived from the expected SSD delay distribution in each bin. The black dashed line represents the one-to-one relation. Similarly, bins where posterior lag distributions showed more than one peak are indicated by stars rather than circles in blue (orange) for ICCF (\textsc{JAVELIN}).}
    \label{fig:error_ssd_obs}
\end{figure}

The remainder of the quoted stacked delay uncertainties outside multi-modal bins can be largely explained by the number of constituent light curve pairs contributing to the estimate. Figure \ref{fig:lag_error_v_N} shows both ICCF and \textsc{JAVELIN} posterior delay semi-amplitudes scale more shallowly than expected for a statistical noise limited system ($\propto1/\sqrt{N}$, where $N$ is the number of objects), although a visual comparison maintains $\propto1/\sqrt{N}$ is still somewhat consistent. The displayed fits do not include the multi-modal measurements (shown by stars) which are seen to have systematically larger uncertainty as expected. While the quoted uncertainties do improve with increased number statistics, they appear bounded by the inherent dispersion in the predicted SSD delay distribution (Figure \ref{fig:error_ssd_obs}) suggesting delay errors are not significantly underestimated. Although, we do note as per our discussion in Section \ref{sec:main_results}, we are likely limited by systematic uncertainties regarding our bin compositions.

\bsp	
\label{lastpage}
\end{document}